\documentclass[11pt,fleqn]{article}

\usepackage{latexsym}
\usepackage{amsmath}
\usepackage{amssymb}
\usepackage{graphicx}


\def\R{ {\mathbb R} }

\def\C{ {\mathbb C} }

\def\M{ {\mathbb M} }
\def\K{ {\mathbb K} }
\def\E{ {\mathbb E} }

\def\re{{\rm e}}

\def\CwedgeX{\mbox{$\bigwedge^{2}({\mathbb C}^{4})$}}
\def\lbk{\lbrack\!\lbrack}
\def\rbk{\rbrack\!\rbrack}

\def\mijnlim#1{\mathop{\rm lim}\limits_{#1}}

\def\bU{{\bf U}}
\def\bV{{\bf V}}

\def\bZ{{\bf Z}}
\def\bA{ {\bf A} }

\def\mi{{\mathrm i}}
\def\me{{\mathrm e}}

\newcommand{\sech}{{\mathop{\rm sech}\nolimits}}
\newcommand{\qmbox}[1]{\quad\mbox{#1}\quad}

\newenvironment{arrayl}%
        {\begin{array}{@{}r@{\hskip\arraycolsep}c@{\hskip\arraycolsep}l}}%
        {\end{array}}

\begin{document}

\title{\textbf{A robust numerical method to study oscillatory instability
    of gap solitary waves}} 
\vspace{.75in}

\author{\textsf{Gianne Derks
\mbox{and} Georg A. Gottwald}\\[6mm]
{\it Dept.\ of Mathematics and Statistics, University of Surrey,
  Guildford, Surrey, GU2 7XH, UK}\\ 
{\it School of Mathematics and Statistics, University of Sydney, NSW
  2006, Australia} 
}
\vspace{.15in}
\maketitle

\vspace{.15in}

\begin{abstract}
\noindent The spectral problem associated with the linearization
about solitary waves of spinor systems or optical coupled mode
equations supporting gap solitons is formulated in terms of the Evans
function, a complex analytic function whose zeros correspond to
eigenvalues. These 
problems may exhibit oscillatory instabilities where
eigenvalues detach from the edges of the continuous spectrum, so
called edge bifurcations. A numerical framework, based on a fast
robust shooting algorithm using exterior algebra is 
described. The complete algorithm is robust in the sense that it does
not produce spurious unstable eigenvalues. The algorithm allows to
locate exactly
where the unstable discrete eigenvalues detach
from the continuous spectrum. Moreover, the
algorithm allows for stable shooting along multi-dimensional stable
and unstable manifolds. The method is illustrated by computing the
stability and instability of gap solitary waves of a coupled mode
model.


\end{abstract}

\newpage


\section{Introduction}\label{sec1}
\setcounter{equation}{0}

Transmitting information efficiently across long optical waveguides is
a big challenge in telecommunications. Gap solitons are potential
candidates to achieve this goal. A gap in the linear spectrum allows
solitons in spinor-like systems to propagate without losing energy
due to a resonant interaction with linear waves
\cite{desterke94,kivshar95}. For example, an optical fiber with a
periodically varying refractive index supports gap solitons. The gap
here is created by Bragg reflection and resonance of waves along the
grating.

Before its application to optical waveguides and transmission of
optical pulses, gap solitons have been studied in the context of
spinor field equations in elementary particle physics
\cite{lee75,lee77} and in condensed matter physics
\cite{campbell82}. There is an extensive literature on the existence of
gap solitons, but the issue of their stability, which is of paramount
practical importance, is still an open question for many systems.

Gap solitons were long believed to be stable. This was conjectured on
the grounds of computer simulations \cite{aceves89} in (restricted)
parameter regimes. Only recently \textsc{Barashenkov, Pelinovsky \&
Zemlyanaya}~\cite{bpz98} showed analytically by using a perturbation
theory, and \textsc{Barashenkov \& Zemlyanaya}~\cite{bz00} verified
numerically that, in a particular optical system, gap solitons can
undergo oscillatory instabilities where eigenvalues detach from the
edges of the continuous spectrum (edge
bifurcations). In~\textsc{Kapitula \& Sandstede}~\cite{kapitula02},
Evans functions are used to detect analytically the onset of an
oscillatory instability (edge bifurcation) in near integrable systems.
From an analytical point of view a difficulty in proving
stability/instability is that the energy in these systems is neither
bounded from above nor from below. Usual techniques such as
energy-momentum methods or methods involving Lyapunov functionals are
therefore bound to fail.

An often used numerical approach to eigenvalue problems is to
discretize the spectral problem on the truncated domain
$x\in[-L_\infty,L_\infty]$ (with $L_\infty \gg 1$) using finite differences,
collocation or spectral methods, reducing it to a very large matrix
eigenvalue problem.  There are two central difficulties with this
approach.  Firstly, in general the exact asymptotic boundary
conditions at $x=\pm L_\infty$ depend on the eigenvalue~$\lambda$ in a
nonlinear way, and so application of the exact asymptotic boundary
conditions changes the problem to a matrix eigenvalue problem, which
is {\it nonlinear in the parameter\/}. So matrix eigenvalue solvers
can no longer be used.  In the above papers~\cite{bpz98,bz00},
artificial boundary conditions such as Dirichlet or periodic boundary
conditions, were applied, in order to retain linearity in the spectral
parameter. 
For a detailed discussion on artificial boundary conditions,
see \textsc{Sandstede \& Scheel}~\cite{sandstede00, sandstede01}.
Secondly, the approximate boundary conditions lead often to
spurious discrete eigenvalues generated from the fractured continuous
spectrum.  If the continuous spectrum is strongly stable (that is, the
continuous spectrum is stable and there is a gap between the
continuous spectrum and the imaginary axis) this does not normally
generate spurious {\it unstable\/} eigenvalues.  However, if the
continuous spectrum lies on the imaginary axis (as often happens with
gap solitons due to the Hamiltonian character of the underlying
system), spurious eigenvalues may be emitted into the unstable half
plane. Indeed, \textsc{Barashenkov \& Zemlyanaya}~\cite{bz00} give an
extreme example, where a large number of spurious unstable eigenvalues
are generated by the matrix discretization (see Figure~1
in~\cite{bz00}).  These problems prohibit their method to be used as a
robust scheme to check stability when no theoretical results are given
to guide the experiments.

\smallskip 
A robust and stable numerical scheme is still an open problem. In this
paper we present a numerical method based on the Evans function and
exterior algebra which does not exhibit spurious unstable
eigenvalues. The Evans function is a complex analytic function whose
zeros correspond to eigenvalues of the spectral problem associated
with the linearization about a solitary wave solution.
The Evans function was first introduced by
\textsc{Evans}~\cite{evans75} and generalized by \textsc{Alexander,
Gardner \& Jones}~\cite{agj90}. To define the Evans function, one
writes the wave equations as a system of first order real equations
with respect to the spatial variable~$x$, such that one gets a system
of the form $\bZ_x=F(\bZ,\bZ_t,\bZ_{tt},\ldots)$.  The study of linear
stability of a solitary wave solution $\widehat{\bZ}(x)$ involves a
linearisation about $\widehat \bZ$ by writing the basic solutions as
$\bZ(x) ={\widehat\bZ}(x)+{\bf u}(x)\,e^{\lambda t}$. This will lead
to a linearized problem of the form
\begin{equation}\label{lse0}
{\bf u}_x = {\bf A}(x,\lambda){\bf u}\,,\quad {\bf u}\in\C^{n},
\end{equation}
where $\lambda\in\C$ is the spectral parameter and ${\bf
A}(x,\lambda)$ is a matrix in $\mathbb{C}^{n\times n}$, whose limit
for $x\to\pm\infty$ exists.  The solitary wave solution ${\widehat
\bZ}$ of a partial differential equation is linearly unstable if for a
spectral parameter $\lambda$ with $\Re(\lambda)>0$ there exists an
associated perturbation ${\bf u}(x)$, which is bounded for all $x$.
An oscillatory instability (edge bifurcation) does happen when the
continuous spectrum is on the imaginary axis and one of the bounded
eigenfunctions of the continuous spectrum develops into an
exponentially decaying eigenfunction with $\Re(\lambda)>0$.

To verify if such bounded eigenfunction~${\bf u}(x)$ exists for a
given value of $\lambda$ with $\Re(\lambda)>0$, it is checked if the
unstable manifold at $x=-\infty$ and the stable manifolds at
$x=\infty$ have a non-trivial intersection. If this is the case, the
solitary wave solution ${\widehat \bZ}$ is unstable with positive
growth rate $\Re(\lambda)>0$. To check the transversality condition
the Evans function is used. Hence the Evans function can be viewed as
a Melnikov function or a Wronskian determinant.

Crucial to the initial construction of the Evans function is the
distribution of the eigenvalues of the `system at infinity', that is,
the matrix ${\bf A}_{\pm\infty}(\lambda)$ which is associated with the
limit as $x\to \pm\infty$ of ${\bf A}(x,\lambda)$.  It is assumed that
no eigenvalues are on the imaginary axis \footnote{The assumption on the
hyperbolicity of ${\bf A}_{\pm\infty}(\lambda)$ can be weakened (see
\cite{gardner98,kapitula98}).} and that the number of
negative eigenvalues is constant for $\lambda\in\Lambda$, where
$\Lambda$ is a simply-connected subset of $\mathbb C$.  Let $k$ be the
number of negative eigenvalues of ${\bf A}_{\pm\infty}(\lambda)$ for
$\lambda\in\Lambda$. Note that in the case of different asymptotic
behaviour at $x=\pm\infty$ as for example in the case of fronts, $k$
can be different at $x=\pm\infty$ in general.

This suggests a naive approach by which one may follow the 
stable/unstable manifolds at $x=\pm \infty$ with a standard shooting
method and check their intersection using the Evans function. This may
be done indeed if the dimension of 
these manifolds is~$1$. Otherwise, any
integration scheme will inevitably just be attracted by the
eigendirection corresponding to the most unstable eigenvalue. 
However, for most systems, the dimensional of the stable/unstable
manifold will be larger than 1. In this paper, we will consider a
model equation with stable/unstable manifolds of dimension~2.  To keep
the eigendirections orthogonal in the course of the numerical
integration, one may employ a Gram-Schmidt orthogonalization method.
However, this is a non-analytic procedure, which will eventually
backfire in the case of oscillatory instabilities where we expect
zero's of the Evans function in the complex plane. Indeed, to locate
those complex eigenvalues, Cauchy's Theorem (argument principle) will
be employed and thus an analytical method is crucial.  In our
numerical method, we will use exterior algebra, which allows for an
analytical calculation of the Evans function.

The numerical method builds on the work in~\cite{ab02,derks02}, where a
numerical algorithm is given for the calculation of the Evans function
and applications are give to the stability and instability in a
fifth-order KdV equation.  In this paper, we will show that a similar
algorithm can be used to determine oscillatory instabilities (edge
bifurcations). The algorithm does not exhibit spurious eigenvalues and
the exact asymptotic boundary conditions are built into the definition
of the Evans function in an analytic way.  The analyticity can then be
utilized to apply Cauchy's principle value theorem to study
stability/instability.  An important feature of the numerical method
is that it involves the use of exterior algebra to describe the system
on a higher dimensional space in which a simple shooting method can be
employed. A similar idea is used in \textsc{Brin}~\cite{brin} and
\textsc{Brin \& Zumbrun}~\cite{bz02} for dealing with instabilities in
viscous fluid flows.

\smallskip
To illustrate our method we will consider the following coupled mode model
\begin{equation}\label{MTM}
\begin{arrayl}
0&=&\mi(u_t+u_x)+v+(|v|^2+\rho|u|^2)u\\
0&=&\mi(v_t-v_x)+u+(|u|^2+\rho|v|^2)v \; .
\end{arrayl}
\end{equation}
This is a model to describe optical pulses in waveguides which have
been grated so the refractive index is varying periodically. In the
case $\rho=0$ this equation is known in field-theory as the {\it
  massive Thirring model} and was shown to be completely integrable
(see for example \cite{aceves89,kaup961}). In a nonlinear optics
context, one has $\rho=1/2$ in periodic Kerr media~\cite{desterke94},
but in other media $\rho$ may range from~0 up to
infinity~\cite{romangoli92}.  The equation~\eqref{MTM} has also been
studied by \textsc{Barashenkov, Pelinovsky \&
  Zemlyanaya}~\cite{bpz98}, \textsc{Barashenkov \&
  Zemlyanaya}~\cite{bz00} and 
in a slightly modified form by
\textsc{Kapitula \& Sandstede}~\cite{kapitula02}. In
~\cite{bpz98,bz00} a heuristic perturbation analysis is used to
analyse the onset of oscillatory instabilities and numerical study is
used to give a more complete picture. The numerical study encountered
serious problems as discussed above. In~\cite{kapitula02}, a relation
between the Evans function and the inverse scattering formalism is
established for integrable systems. This forms the basis of a rigorous
perturbation analysis for 
perturbations of the massive Thirring model.

The coupled mode model~\eqref{MTM} is chosen for 
illustration purposes, since
the work in the previous papers allows us to illustrate the advantages
of our method. We stress though that the method we present is general
and can be applied to gap solitons in other systems as well. Moreover,
there is no need for the system to be related to an integrable system.


\section{The numerical Evans function for the coupled mode mode}
\label{sec3}
\setcounter{equation}{0}

The solutions and dynamics of~\eqref{MTM} are best described by
splitting off the real and imaginary part of the fields $u,v$. We
shall write for a solution $u=Q_1+iP_1$ and $v=Q_2+iP_2$ and 
collect the information in a single real solution vector ${\bf
{Z}}=(Q_1,Q_2,P_1,P_2)$.

\subsection{The model equations and its solutions}
We introduce the Lorentz transformation $X=(x-Vt)/\sqrt{1-V^2}$ and
$T=(t-Vx)/\sqrt{1-V^2}$. In these boosted variables the model
system~(\ref{MTM}) may be written in a (quasi-)multisymplectic
framework as
\begin{equation}\label{MTMMS0}
\E[\M{\bf Z}_T+\K{\bf Z}_X]=\nabla S({\bf Z})\,
\end{equation}
where $\bZ=(Q_1,Q_2,P_1,P_2)$, 
\[
S(Z)= \frac12\left[Q_1Q_2+P_1P_2+(Q_1^2+P_1^2)(Q_2^2+P_2^2)\right] +
\frac\rho2\left[(Q_1^2+P_1^2)^2+(Q_2^2+P_2^2)^2\right],
\] 
%
\[
\E=
\begin{pmatrix}
\E_1&0\\0&\E_1
\end{pmatrix},
\qquad
\E_1=
\begin{pmatrix}
e^{-y} & 0 \\
0 & e^y
\end{pmatrix},
\qquad
\M=
\begin{pmatrix}
0 & \sigma_0\\
-\sigma_0 & 0
\end{pmatrix}
\quad {\rm and} \quad
\K=
\begin{pmatrix}
0 & \sigma_3\\
-\sigma_3 & 0
\end{pmatrix},
\]
with $V=\tanh{y}$ and the Pauli matrices $\sigma_{0,3}$ are defined as
\[
\sigma_0=
\begin{pmatrix}
  1&0\\0&1
\end{pmatrix} \quad {\rm and} \quad
\sigma_3=
\begin{pmatrix}
  1&0\\0&-1
\end{pmatrix}. \qquad
\]
Note that if $\rho=0$, the system is invariant under this Lorentz
transformation. 

%
%
The reason for introducing this formalism is that a multisymplectic
formulation allows for a systematic linearization and also sheds light
on conservation properties within the system. We note
that~(\ref{MTMMS0}) is equivariant under action of the continuous
symmetry group $SO(2)$ acting on $\R^4$ represented by
\begin{eqnarray}\label{G}
G_\psi=
\begin{pmatrix}
\cos(\psi)\,\sigma_0 & -\sin(\psi)\sigma_0\\
\sin(\psi)\,\sigma_0 &  \cos(\psi)\,\sigma_0
\end{pmatrix}
\; ,
\end{eqnarray}
since $[G_\psi,\M]=[G_\psi,\K]=[G_\psi,\E]=0$ and $S(G_\psi{\bf
Z})=S({\bf Z})$ for any $\psi$. According to Noether's Theorem, there
is a conservation law associated with this continuous symmetry, namely
$P_T+Q_X=0$ with $P$ and $Q$ determined by
\begin{eqnarray}\label{PQ}
\nabla P({\bf Z})=\M\left.\frac{d}{d\psi}\right|_{\psi=0}
G_\psi(\bZ) 
\quad {\rm and} \quad
\nabla Q({\bf Z})=\K\left.\frac{d}{d\psi}\right|_{\psi=0}
G_\psi(\bZ) \;, 
\end{eqnarray}
hence 
\[
P(\bZ)=\sum_{i=1}^2 Q_i^2+P_i^2 \qmbox{and}
Q(\bZ)=\sum_{i=1}^2 (-1)^{i+1}(Q_i^2+P_i^2).
\]
Note that in the original system~\eqref{MTM}, this symmetry shows up
as an equivariance of the system under simultaneous phaseshifts of $u$
and $v$, i.e., $u\mapsto u\,e^{\mi\psi}$ and $v\mapsto
v\,e^{\mi\psi}$.

Going to a frame moving with the symmetry group and writing the
solutions as ${\bZ}(X,T)=G_{\varphi(X)-\Omega T}\widetilde{\bZ}(X,T)$,
the equations~\eqref{MTMMS0} become
\begin{eqnarray}\label{MTMMS}
\E[\M{\bf Z}_T+\K{\bf Z}_X-\Omega \nabla P ({\bf Z}) 
+\varphi_X\nabla Q({\bf Z})] = \nabla S({\bf Z})\, ,
\end{eqnarray}
where we dropped the tildes.
Time independent solutions in the moving frame were found by
\textsc{Aceves \& Wabnitz}~\cite{aceves89} to be
\begin{eqnarray}\label{solution}
\hat {\bf Z}(X)=\alpha\E^{-\frac12} 
\begin{pmatrix}\sigma_3 & 0\\
0 & \sigma_0
\end{pmatrix}
\begin{pmatrix}
W_r(X)\\
W_r(X)\\
W_i(X)\\
W_i(X)
\end{pmatrix}
\; ,
\end{eqnarray}
where 
\[
\Omega=\cos{\theta}\; (0<\theta<\pi),\quad
\varphi(X)=2\alpha^2\rho\sinh(2y)\arctan
\left\{\tanh[(\sin(\theta))X]\tan{\textstyle\frac{\theta}{2}}\right\},\quad
\]
\[
\alpha=\frac1{\sqrt{1+\rho\cosh{2y}}}\, \qmbox{and}
W(X)=W_r(X)+\mi W_i(X)=
\frac{\sin(\theta)}{\cosh{((\sin{\theta})X-\mi\theta/2)}}.
\]
Note that there cannot be any gap solitons for $|\Omega|>1$ as then
there is no gap in the linear spectrum.  At the upper edge of the gap,
at $\theta\rightarrow 0$
%
($\Omega\to1$), the solutions approaches the small-amplitude
nonlinear Schr\"odinger soliton
$W(X)=\theta\sech(\theta(X-\mi/2))$. At the lower edge, at $\theta
\rightarrow \pi$
%
($\Omega\to-1$), the gap soliton has a finite amplitude and decays
algebraically, $W(X)=\mi/(X+\mi/2)$. These two limits are referred to as
``low intensity'' and ``high intensity'' limits, respectively
\cite{desterke94}.


\subsection{The linearized problem}

To study the stability of the solution (\ref{solution}), we linearize
around this solution in the moving frame and use a spectral Ansatz
\begin{equation}\nonumber
{\bf Z}(X,T)=
\left({\bf \widehat Z}(X)+{\bf u}(X)e^{\lambda T}\right).
\end{equation}
We obtain the dynamical system
\begin{equation}\label{lse}
{\bf u}_X = {\bf A}(X,\lambda){\bf u}\,,\quad {\bf u }\in\C^{4}\; ,
\end{equation}
with
\begin{equation}\label{Axl}
{\bf A}(X,\lambda)
=
\K^{-1}[
\E^{-1}D^2S({\bf \hat Z})
+\Omega D^2P({\bf \hat Z})
-\varphi_XD^2Q({\bf \hat Z})
-\lambda \M]
\end{equation}
where $D^2$ denotes the Hessian.

\subsection{Asymptotic properties of the linearised system}\label{sec.as}
As described in the Introduction, the eigenvalues and eigenfunctions
of the asymptotic matrix
$\bA_\infty(\lambda)=\mijnlim{x\to\pm\infty}{\bf A}(x,\lambda)$ are
crucial for the numerical Evans function approach. The matrix ${\bf
A}(x,\lambda)$ has the asymptotic property that
\begin{equation}\label{Ainfty}
{\bf A}_\infty(\lambda) = 
\lim_{x\to\pm\infty}{\bf A}(x,\lambda)=
\begin{pmatrix}
-\lambda & 0 & -\Omega & -e^y \\
0 & \lambda & e^{-y} & \Omega \\
\Omega & e^{y} & -\lambda & 0 \\
-e^{-y} & -\Omega & 0 & \lambda
\end{pmatrix}\;.
\end{equation}
The characteristic polynomial of ${\bf A}_\infty(\lambda)$ is
\begin{equation}\label{cpoly}
\Delta(\mu,\lambda) = {\rm det}[
\mu{\bf I}-{\bf A}_\infty(\lambda)] =
(\mu^2-\lambda^2-1+\Omega^2)^2+4\Omega^2\lambda^2\;.
\end{equation}
Thus the eigenvalues $\mu$ of the asymptotic matrix satisfy
\begin{equation}\label{eq.eigval}
\mu^2-\lambda^2-1+\Omega^2 = \pm 2\mi \Omega\lambda
\qmbox{or}
\mu^2= 1+(\lambda\pm\mi\Omega)^2\;.
\end{equation}

The continuous spectrum is found by setting ${\cal {R}}(\mu)=0$ or
$\mu=\mi k$. A short calculation gives that there are four branches of
continuous spectrum on the imaginary $\lambda$-axis:
\[
\lambda_\pm(k) = \mi(\sqrt{1+k^2} \pm |\Omega|) \qmbox{and}
-\lambda_\pm(k), \quad k\in\mathbb{R} \; .
\]
The end points of the continuous spectrum are at $\pm\mi(1-|\Omega|)$
and $\pm\mi(1+|\Omega|)$. For all values of $\lambda$ not in the
continuous spectrum, there are 2 eigenvalues~$\mu$ with positive real
part and 2 eigenvalues~$\mu$ with negative real part. Thus outside the
continuous spectrum, there is a $2:2$ splitting, that is the dimension
of the stable/unstable manifolds is $2$ at both limits of
$x=\pm\infty$.

With~\eqref{eq.eigval}, it is easy to determine explicit expressions
for the eigenvalues and eigenvectors. The eigenvalues are
\[
\begin{arrayl}
\mu_m^\pm(\lambda) &=& \pm\sqrt{(\lambda-\mi|\Omega|)^2+1} = 
\pm \sqrt{\lambda-\mi(|\Omega|+1)}\,\sqrt{\lambda-\mi(|\Omega|-1)},\\
\mu_p^\pm (\lambda)&=& \pm\sqrt{(\lambda+\mi|\Omega|)^2+1} = 
\pm \sqrt{\lambda+\mi(|\Omega|+1)}\,\sqrt{\lambda+\mi(|\Omega|-1)}. 
\end{arrayl}
\]
where the square roots are defined as follows:
\[
\sqrt{z} = \sqrt{|z|}\,
\me^{\mi\arg(z)/2}, \qmbox{where}
-\pi<\arg(z)\leq \pi \qmbox{and}  z= \lambda\pm\mi(|\Omega|\pm1).
\]
Hence the cuts in the complex plane, associated with this definition of
the square roots, start at the end points of the continuous spectrum
and continue in the left half of the complex $\lambda$-plane. Note
that this definition implies that all square roots have positive real
parts.

In this model system it is easy to find explicit expressions for the
eigenvectors.  The eigenvector with the eigenvalue~$\mu^\pm_i$ for the
matrix~$\bA_\infty(\lambda)$ is given by
\[
\begin{arrayl}
\mbox{\bf vec}_m^\pm (\lambda)&=& 
(F^\pm_m,\mbox{sgn}(\Omega) \mi e^{-y}, 
-\mbox{sgn}(\Omega) \mi F^\pm_m , e^{-y} )^T,
\qmbox{where} F^\pm_m = \lambda-\mu_m^\pm  - \mi|\Omega| ,\\
\mbox{\bf vec}_p^\pm (\lambda)&=& 
(F^\pm_p,-\mbox{sgn}(\Omega) \mi e^{-y}, 
\mbox{sgn}(\Omega) \mi F^\pm_p , e^{-y} )^T,
\qmbox{where} F^\pm_p = \lambda-\mu_p^\pm  + \mi|\Omega| .
\end{arrayl}
\]
The eigenvector with the eigenvalue~$\overline{\mu^\pm_i}$ for the
adjoint matrix~$(\overline{\bA_\infty(\lambda)})^T$ is given by
\[
\begin{arrayl}
\mbox{\bf advec}_m^\pm (\lambda)&=& 
(\overline{F^\pm_m},\mbox{sgn}(\Omega) \mi e^{y}, 
-\mbox{sgn}(\Omega) \mi \overline{F^\pm_m} , e^{y} )^T,
\qmbox{where} F^\pm_m = \lambda-\mu_m^\pm  - \mi|\Omega| ,\\
\mbox{\bf advec}_p^\pm (\lambda)&=& 
(\overline{F^\pm_p},-\mbox{sgn}(\Omega) \mi e^{y}, 
\mbox{sgn}(\Omega) \mi \overline{F^\pm_p} , e^y )^T,
\qmbox{where} F^\pm_p = \lambda-\mu_p^\pm  + \mi|\Omega| .
\end{arrayl}
\]

One might also use a numerical routine to solve the eigenvalue problem
of ${\bf A}_\infty(\lambda)$. However, one has to be very careful near
the end points of the continuous spectrum. At these points two
eigenvalues collide, hence it will be tricky to find numerically the
'correct' eigenvalue. If we fail to do so, this will result in a
change of orientation in the Evans function~(\ref{2.9}). This causes
unwanted zero's of the Evans function and is reminiscent of the
unstable spurious eigenvalues observed when the system~(\ref{lse}) is
solved by some form of discretization as done by \textsc{Barashenkov
\& Zemlyanaya}~\cite{bz00}.  One can construct a numerical routine to
follow the correct eigenfunction, but if the analytical expressions
are readily available (as they are here) it is much easier.

\subsection{The Evans function and the formulation in the exterior
  algebra}\label{sec.evans}

For $\Re(\lambda)>0$, the system~(\ref{lse}) and the properties of the
system at infinity, ${\bf A}_\infty(\lambda)$, are in standard form
for the dynamical systems formulation of the spectral problem proposed
by \textsc{Evans}~\cite{evans75} and generalized by \textsc{Alexander,
Gardner \& Jones}~\cite{agj90}.  
A value of $\lambda\in\Lambda$ is an eigenvalue if 
the $2$-dimensional space of solutions which decays as
$x\to-\infty$ and the the $(4-2)$-dimensional space of solutions which
do not grow exponentially as $x\to+\infty$
have a nontrivial intersection. The {\it Evans function\/} is an
analytic function which gives a
zero if such a nontrivial intersection exists.
To obtain an analytic description of the $2$-dimensional space of
solutions of (\ref{lse}) which do not grow exponentially as
$x\to+\infty$, we will use that the system~(\ref{lse}) induces a
dynamical system on the wedge-space~$\CwedgeX$. 
This is a space of dimension ${4\choose 2}=6$.
To define the Evans function, the induced dynamics on this
wedge-space~$\CwedgeX$ will be used.  

The induced system can be written as
\begin{equation}\label{Unk}
{\bf U}_x = {\bf A}^{(2)} (x){\bf U} , \quad {\bf U}\in\CwedgeX.
\end{equation}
Here the linear operator (matrix) ${\bf A}^{(2)}$ is defined on a
decomposable $2$-form~${\bf u}_1\wedge{\bf u}_2$, $ {\bf
u}_i\in\C^{\,4}$, as
\begin{equation}\label{eq:defA2}
{\bf A}^{(2)} ({\bf u}_1\wedge{\bf u}_2) := 
({\bf A} {\bf u}_1)\wedge{\bf u}_2 +
{\bf u}_1\wedge({\bf A} {\bf u}_2)\,
\end{equation}
and it extends by linearity to the non-decomposable elements in
$\CwedgeX$.  This construction can be carried out in a coordinate free
way and can be generalised to $(k)$:$(n-k)$ splittings in
$n$-dimensional dynamical systems. General aspects of the numerical
implementation of this theory can be found in \textsc{Allen \&
Bridges}~\cite{ab02}.

Since the induced matrix ${\bf A}^{(2)}(x,\lambda)$ inherits the
differentiability and analyticity of ${\bf A}(x,\lambda)$, the
limiting matrices will exist,
\[
{\bf A}_\infty^{(2)}(\lambda) = 
\lim_{x\to\pm\infty}{\bf A}^{(2)}(x,\lambda)\,.
\]
The set of eigenvalues of the matrix ${\bf A}_\infty^{(2)}(\lambda)$
consists of all possible sums of 2~eigenvalues of ${\bf
A}_\infty(\lambda)$ (this is an exercise in multi-linear algebra, see
\textsc{Marcus}~\cite{marcus}).  Therefore, for $\Re(\lambda)>0$,
there is an eigenvalue of ${\bf A}_\infty^{(2)}(\lambda)$, denoted by
$\sigma_+(\lambda)$, which is the sum of the $2$ eigenvalues of ${\bf
A}_\infty(\lambda)$ with negative real part, i.e., 
%
$\sigma_+(\lambda)= -(\mu_m^+(\lambda)+\mu_p^+(\lambda))$
(note that the subscript ``$+$'' in $\sigma_+(\lambda)$ refers to exponentially
decaying behaviour at $+\infty$).
Moreover this eigenvalue is
simple, an analytic function of $\lambda$ and has real part strictly
less than any other eigenvalue of ${\bf A}_\infty^{(k)}(\lambda)$.
Similarly, 
%
there is an eigenvalue
$\sigma_-(\lambda)$, which is the sum of the $2$ eigenvalues of ${\bf
A}_\infty(\lambda)$ with non-negative real part, i.e.,
$\sigma_-(\lambda) = \mu_m^-(\lambda)+\mu_p^-(\lambda)$, and
$\sigma_-(\lambda)$ is simple, an analytic function of $\lambda$, and
has real part strictly greater than any other eigenvalue of ${\bf
A}_\infty^{(2)}(\lambda)$. Note that
$\sigma_-(\lambda)=-\sigma_+(\lambda)$ in this example.

Let $\zeta^\pm(\lambda)$ be the eigenvectors associated with
$\sigma_\pm(\lambda)$, defined by
\begin{equation}\label{eigVs}
{\bf A}_\infty^{(2)}(\lambda)\zeta^+(\lambda) =
\sigma_+(\lambda)\zeta^+(\lambda)
\qmbox{and}
{\bf A}_\infty^{(2)}(\lambda)\zeta^-(\lambda) =
\sigma_-(\lambda)\zeta^-(\lambda)\,.
\end{equation}
These vectors can always be constructed in an analytic way (see
\cite{derks02}). From section~\ref{sec.as}, it follows that 
\[
\zeta^\pm(\lambda) = \mbox{\bf vec}_m^\pm (\lambda) \wedge 
\mbox{\bf vec}_p^\pm (\lambda).
\]
This implies
\[
\zeta^\pm= 
\left(
F^\pm_p+F^\pm_m, -2 e^{y} F^\pm_p F^\pm_m, 
\mbox{sgn}(\Omega)\mi(F^\pm_m-F^\pm_p),
\mbox{sgn}(\Omega)\mi(F^\pm_m-F^\pm_p), -2 e^{-y} , F^\pm_p+F^\pm_m 
\right)^T ,
\]
The solution $\bU^\pm(x,\lambda)$ is the solution of the linearised
system~\eqref{Unk} with the property that 
$\lim_{x\to\pm\infty} e^{-\sigma_\pm(\lambda)x} \bU^\pm(x,\lambda)
= \zeta^\pm(\lambda)$. 
In
this example, dimension of the unstable manifolds at $x=\infty$ and
$x=-\infty$ are the same. For the construction of the general case
see~\cite{ab02,derks02}.

Note that the solutions ${\bf U}^\pm(x,\lambda)$ are analytic
expressions which represent the space of solutions which decay as
$x\to\pm\infty$ in the original system~(\ref{lse}). With this the
Evans function can be defined as 
\begin{equation}\label{2.9}
E(\lambda) = \re^{-\int_0^x\tau(s,\lambda)ds}\, {\bf U}^-(x,\lambda)\wedge
{\bf U}^+(x,\lambda)\,,\quad \lambda\in\Lambda\,,
\end{equation}
where $\wedge$ is the wedge product and
\begin{equation}\label{2.10}
\tau(x,\lambda) = {\rm Tr}( {\bf A}(x,\lambda) )\,.
\end{equation}
For the case of the coupled mode equation~\eqref{MTM} this expression
simplifies, since ${\rm Tr}( {\bf A}(x,\lambda) )=0$, see~(\ref{Axl}).
The Evans function, as defined above, can be extended across the
continuous spectrum with some cuts in the complex plane,
see~\textsc{Gardner \& Zumbrun}~\cite{gardner98} and \textsc{Kapitula
\& Sandstede}~\cite{kapitula98}.

Next we will give an equivalent description of the Evans function,
using the adjoint system. The adjoint system of (\ref{Unk}) is
\begin{equation}\label{Vminus}
{\bf V}_x = -\overline{[{\bf A}^{(2)}(x)]}^T{\bf V}\,.
\end{equation}
The dimension of the unstable manifold of this adjoint system is equal
to the dimension of the stable manifold of the linearised
system~\eqref{Unk} and its most unstable eigenvalue
is~$-\overline{\sigma^+}$. Using the Hodge-star operator, we can
relate the most unstable solution $\bV^-$ at $x=-\infty$ of the
adjoint system with the most unstable solution~$\bU^-$ of the
linearised system at $x=-\infty$. Details can be found
in~\cite{ab02,bsz01,derks02}. To formulate the alternative description
of the Evans function, we have to construct an inner product on
$\CwedgeX$. Let
\[
{\bf U} = {\bf u}_1\wedge{\bf u}_2
\quad{\rm and}\quad{\bf V} =
{\bf v}_1\wedge{\bf v}_2\,,\quad
{\bf u}_i,{\bf v}_j\in\C^{\,4}\,,\quad\forall\ i,j=1,2\,,
\]
be any decomposable $2$-forms.  The inner product of
${\bf U}$ and ${\bf V}$ is defined by
\[
\lbk {\bf U} , {\bf V} \rbk_2 := {\rm det}\left[
\begin{matrix}
\langle {\bf u}_1,{\bf v}_1\rangle & \langle{\bf u}_1,{\bf v}_2\rangle\\
\langle {\bf u}_2,{\bf v}_1\rangle & \langle{\bf u}_2,{\bf v}_2\rangle
\end{matrix}\right]\,,
\]
where $\langle\cdot,\cdot\rangle_4$ is the complex inner product in
$\mathbb{C}^{\,4}$. Since every element in $\CwedgeX$ is a sum of
decomposable elements, this definition extends by linearity to any
$2$-form in $\CwedgeX$.
An equivalent definition of the Evans function~\eqref{2.9} is given by
the following readily computable expression
\begin{equation}\label{UVmatch1}
E(\lambda) = \lbk{{\bf V}^-(0,\lambda)},
{\bf U}^+(0,\lambda)\rbk_2\,,
\end{equation}
where $\bV^-(x,\lambda)$ is the most unstable solution at $x=-\infty$
of the adjoint system~\eqref{Vminus}. From Section~\ref{sec.as}, it
follows that the eigenvector of the adjoint
matrix~$-\overline{\left(\bA^{(2)}\right)}^T$ for the eigenvalue
$-\overline{\sigma_+}=\overline{\mu_p^+ + \mu_m^+}$ is
\[
\eta^-=
\left(
\overline{F^+_p+F^+_m}, -2 e^{y} \overline{F^+_p F^+_m}, 
\mbox{sgn}(\Omega)\mi(\overline{F^+_m-F^+_p}),
\mbox{sgn}(\Omega)\mi(\overline{F^+_m-F^+_p}), 
-2 e^{-y} , \overline{F^+_p+F^+_m} 
\right)^T .
\]
Hence the solution $\bV^-(x,\lambda)$ is the solution of the adjoint
linearised system~\eqref{Vminus} with the asymptotic behaviour
$\mijnlim{x\to-\infty} e^{\overline{\sigma_+(\lambda)}x}
\bV^-(x,\lambda) = \eta^-(\lambda)$. The generalisation of this
definition for general splittings and more details can be found
in~\cite{bsz01,derks02}.

For the numerical implementation, we will need a basis for $\CwedgeX$,
and the above construction assures that any basis will do.  Therefore
there is no loss of generality in assuming that the bases chosen are
the standard ones.  Starting with the standard basis for $\C^{4}$, and
volume form ${\cal V} = {\bf e}_1\wedge\cdots\wedge{\bf e}_4$, let
${\bf a}_1,\ldots,{\bf a}_{6}$ be the induced orthonormal basis on
$\CwedgeX$.  Using a standard lexical ordering, this basis can be
taken to be
\begin{equation}\label{5.1}
\begin{array}{l}
{\bf a}_1 = {\bf e}_1\wedge{\bf e}_2 ,\quad
{\bf a}_2 = {\bf e}_1\wedge{\bf e}_3 ,\quad
{\bf a}_3 = {\bf e}_1\wedge{\bf e}_4 ,\\
{\bf a}_4 = {\bf e}_2\wedge{\bf e}_3 ,\quad
{\bf a}_5 = {\bf e}_2\wedge{\bf e}_4 ,\quad
{\bf a}_6 = {\bf e}_3\wedge{\bf e}_4 .
\end{array}
\end{equation}
Any ${\bf U}\in\CwedgeX$ can be expressed as ${\bf U}=
\sum_{j=1}^{6}U_j{\bf a}_j$. Since the basis elements ${\bf a}_i$ are
orthogonal and the inner product $\lbk\cdot,\cdot\rbk_2$ on $\CwedgeX$
is equivalent to the inner product $\langle\cdot,\cdot\rangle_{6}$ on
$\mathbb{C}^{6}$, the expression~\eqref{UVmatch1} for the Evans
function can be expressed in the equivalent form
\begin{equation}\label{UVmatch}
E(\lambda) = \langle {{\bf V}^-(0,\lambda)},
{\bf U}^+(0,\lambda)\rangle_{6}\,.
\end{equation}
It is this form of the Evans function we implement in our numerical
algorithm.

The matrix ${\bf A}^{(2)}:\CwedgeX\to\CwedgeX$ can be associated with
a complex $6\times 6$ matrix with entries
\begin{equation}\label{entries1}
\{ {\bf A}^{(2)}\}_{i,j} = \lbk {\bf a}_i,{\bf
  A}^{(2)}{\bf a}_j\rbk_2\,,\quad i,j=1,\ldots,6\,,
\end{equation}
where, for any decomposable ${\bf U}={\bf u}_1\wedge{\bf
u}_2\in\CwedgeX$, ${\bf A}^{(2)}{\bf U} := {\bf A}{\bf u}_1\wedge{\bf
u}_2 + {\bf u}_1\wedge{\bf A}{\bf u}_2$.  Let ${\bf A}$ be an
arbitrary $4\times 4$ matrix with complex entries $a_{ij},
i,j=1,\cdots,4$, then, with respect to the basis (\ref{5.1}), ${\bf
A}^{(2)}$ takes the explicit form
\[
{\bf A}^{(2)}=
{\small
\left[\begin{matrix}
a_{11}\!+\! a_{22} & a_{23} & a_{24} & -a_{13} & -a_{14} & 0\\
a_{32} & a_{11}\!+\! a_{33} & a_{34} & a_{12} & 0 & -a_{14}\\
a_{42} & a_{43} & a_{11}\!+\! a_{44} & 0 & a_{12} & a_{13}\\
-a_{31}& a_{21}& 0 & a_{22} \!+\! a_{33} & a_{34} & -a_{24}\\
-a_{41}& 0 & a_{21} & a_{43} & a_{22} \!+\! a_{44} & a_{23}\\
0  & -a_{41}& a_{31}& -a_{42} & a_{32} & a_{33} \!+\! a_{44}
\end{matrix}\right]
}
\]
Details for these constructions in more general systems can be found
in~\cite{ab02}.

\subsection{Integration scheme}

The subtle nature of the oscillatory instability requires a high order
integration scheme for the numerical integration of the linearised and
the adjoint system. Instead of using the second order Gauss-Legendre
Runge-Kutta method, i.e. the implicit midpoint rule, as in
\textsc{Bridges, Derks \& Gottwald}\cite{derks02}, we employ here a
fourth order Gauss-Legendre scheme.
We solve in ${\mathbb C}^6$
\begin{eqnarray}\label{GL1}
{\bf U}^{n+1}={\bf U}^n +\frac{1}{2}\Delta x({\bf K_1}+{\bf K_2}),
\end{eqnarray}
where $\Delta x$ is the spatial step size, sub- and superscripts $n$
denote the spatial discretization and ${\bf K_{1,2}}$ are implicitly
defined by
\begin{eqnarray}
{\bf K_1}&=&{\bf A}^{(2)}(x_n+(\frac{1}{2}+\frac{\sqrt{3}}{6})\Delta x)
\times
({\bf U}^n + \frac{1}{4}\Delta x {\bf K_1} +
(\frac{1}{4}+\frac{\sqrt{3}}{6})\Delta x {\bf K_2}) \nonumber \\
\label{GL2}
{\bf K_2}&=&{\bf A}^{(2)}(x_n+(\frac{1}{2}-\frac{\sqrt{3}}{6})\Delta x)
\times
({\bf U}^n + \frac{1}{4}\Delta x {\bf K_2} +
(\frac{1}{4}-\frac{\sqrt{3}}{6})\Delta x {\bf K_1})\; .
\end{eqnarray}
In practice we solve (\ref{GL2}) for ${\bf K_{1,2}}$ and then
subsequently we can solve (\ref{GL1}) for $\bf U^{n+1}$. Since all
equations are linear, the implicit form of (\ref{GL2}) can be cast in
an explicit form.

The procedure for the numerical calculations is as follows. As
explained in Section~\ref{sec.evans}, it is sufficient to restrict the
shooting algorithm to $\CwedgeX$. As a starting vector for the
shooting algorithm we use the eigenvectors of ${\bf
  A}_\infty^{(2)}(\lambda)$ in the far-field (see
Sections~\ref{sec.as} and~\ref{sec.evans}). For the integration of the
linearised system starting at $x=+L_\infty$ (with $L_\infty\gg1$) the
starting vectors for each $\lambda$ are the eigenvectors
$\zeta^+(\lambda)$ related to the eigenvalues with the largest
negative real part; for the integration of the adjoint system starting
at $x=-L_\infty$ the starting vectors are the eigenvectors
$\eta^-(\lambda)$ related to the eigenvalues with the largest positive
real part. We have build in an analytical normalisation process such
that the eigenvectors are normalized so that
$\langle{\eta^-(\lambda)},\zeta^+(\lambda)\rangle_{6} =1$, for large
values of $\lambda$. (Note that this normalisation is not used for
values of $\lambda<2$.)

To calculate the Evans function, the linearised equation on $\CwedgeX$
\begin{equation}\label{xplus}
\frac{d\ }{dx}\widetilde{\bf U}^+ 
= [{\bf A}^{(2)}(x,\lambda) - \sigma_+(\lambda){\bf I}_d]
\widetilde{\bf U}^+\,,
\quad \widetilde{\bf U}^+(x,\lambda)\big|_{x=L_\infty} =
\zeta^+(\lambda)\,,
\end{equation}
is integrated from $x=L_\infty$ to $x=0$, where the scaling
\begin{equation}\label{scalep}
\widetilde{\bf U}^+(x,\lambda) = e^{-\sigma_+(\lambda)x}\,
{\bf U}^+(x,\lambda)
\end{equation}
ensures that any numerical errors due to the exponential growth are
removed and $\widetilde{\bf U}^+(x,\lambda)\big|_{x=0}={\bf
U}^+(x,\lambda)\big|_{x=0}$ is bounded.  An alternative to this
scaling is to impose a renormalization of the vectors during or at the
end of the integration, with for example $|\widetilde{\bf
U}^+(0,\lambda)|=1$, but such a scaling does not preserve analyticity.

For $x<0$, the adjoint equation 
\begin{equation}\label{xminus}
\frac{d\ }{dx}\widetilde{\bf V}^- 
= [ -\overline{{\bf A}^{(2)}(x,\lambda)}^T  +
\overline{\sigma_+(\lambda)} {\bf I}_d ]
\widetilde{\bf V}^-\,,
\quad \widetilde{\bf V}^-(x,\lambda)\big|_{x=-L_\infty} =
{\eta^-(\lambda)}\,,
\end{equation}
is integrated from $x=-L_\infty$ to $x=0$, also using the implicit
midpoint rule, where again we introduce a rescaling 
\begin{equation}\label{scalem}
\widetilde{\bf V}^-(x,\lambda) = e^{\overline{\sigma_+(\lambda)}x}\,
{\bf V}^-(x,\lambda)\
\end{equation}
to remove the exponential growth. 

At $x=0$, the computed Evans function is
\begin{equation}\label{enumeric}
E(\lambda) = \langle {{\bf V}^-(0,\lambda)},
{\bf U}^+(0,\lambda)\rangle_{6} =
\langle {\widetilde{\bf V}^-(0,\lambda)},
\widetilde{\bf U}^+(0,\lambda)\rangle_{6} \,.
\end{equation}


\section{Numerical results and discussion}

\label{sec4}
\begin{figure}[htb]
  \includegraphics[width=.95\textwidth]{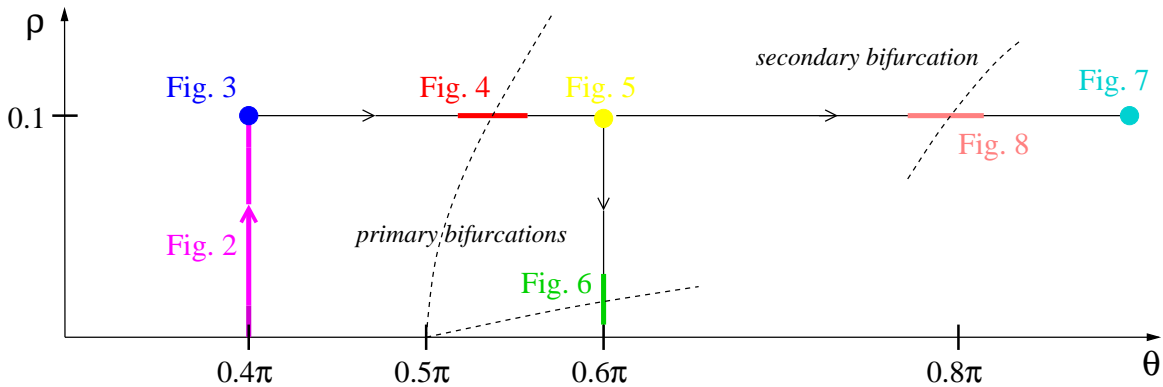}
  \caption{Cartoon to illustrate where in parameter space the
    numerical analysis takes place.}
  \label{fig0}
\end{figure}

\setcounter{equation}{0}
In this Section we show results of our algorithm for the detection of
oscillatory instabilities in the perturbed massive Thirring model
(\ref{MTM}). We do not attempt here to present a thorough numerical
analysis of the bifurcation scenarios of (\ref{MTM}). The reader is
referred to \cite{bz00}. Instead our objective here is the presentation
of a numerical algorithm which uses the Evans function as a
numerical diagnostic tool for analysing edge bifurcations. Therefore
we limit ourselves to illustrating several features of our numerical method.
In Figure~\ref{fig0}, we have sketched a cartoon which indicates
where in parameter space, i.e., in the $\theta$-$\rho$-plane, the
numerical analysis takes place.
This cartoon is a guidance to help to locate the upcoming plethora of
figures and how these figures are related to bifurcations and
instabilities.  

We study instability by computing the Evans function $E(\lambda)$ as
defined in (\ref{UVmatch}). The analyticity of the Evans function for
$\Re(\lambda)>0$ allows one to detect oscillatory instabilities, i.e.
complex roots of the Evans function, by means of Cauchy's theorem. 
The winding number of a closed curve in the $\lambda$-plane tells us about
the number of unstable eigenvalues. 
In all our calculations, we compute the complex Evans
function~$E(\lambda)$, while varying the spectral
parameter~$\lambda=\mi\widetilde \lambda$ on the imaginary axis or
varying the spectral parameter parallel to the imaginary axis with a
(small) offset, explicitly $\lambda=\mbox{off} + \mi \widetilde
\lambda$, where ``{off}'' is the offset.  
The normalisation of the Evans function is chosen such that the Evans
function converges to~1 for $\lambda$ large. Hence the closed curve is
formed by connecting the endpoints of the imaginary axis via the
half-circle with infinite radius. On this half-circle, the Evans function
will always have the value~1.

Since the system~(\ref{MTM}) has translational and rotational
symmetry, the Evans function will have a fourth order zero at
$\lambda=0$. This means that in the vicinity of $\lambda=0$, the Evans
function scales as $E(\lambda)\sim
\lambda^4$. Hence even an offset of only $10^{-4}$ yields that
$E(\lambda)$ is of the order of $10^{-16}$, making the calculations
meaningless. In order to avoid this problem, we use an offset of at
least $5\cdot 10^{-3}$ to analyse the Evans function near $\lambda=0$.
Since we are interested here in oscillatory instabilities which occur
at the edges of or within the continuous spectrum, and not in
translational instabilities where eigenvalues emanate from
$\lambda=0$, the offset near the $\lambda=0$ does not affect the
detection of instabilities.

\begin{figure}[htb]
\,
\includegraphics[width=.47\textwidth]{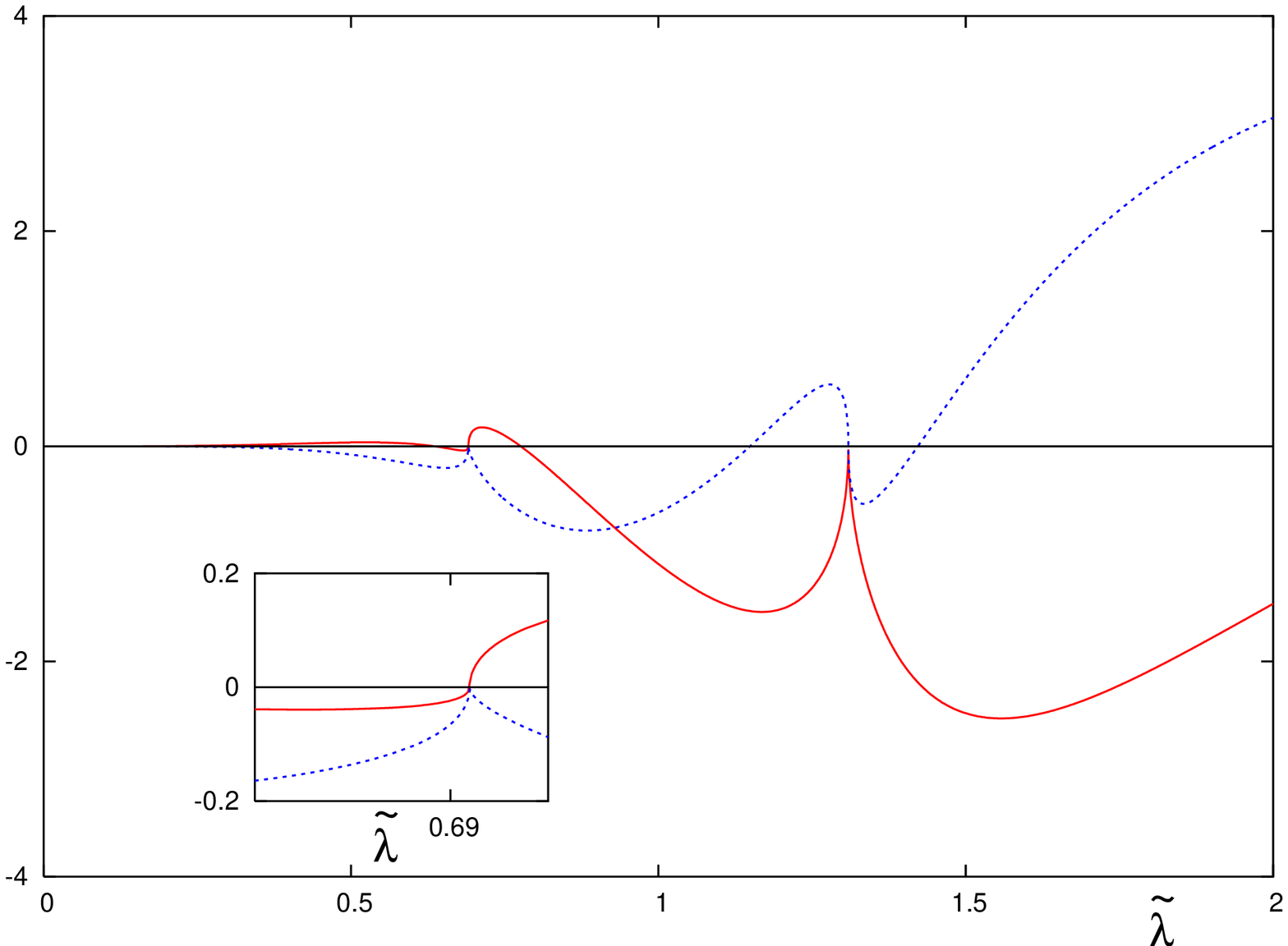}
\qquad
\includegraphics[width=.47\textwidth]{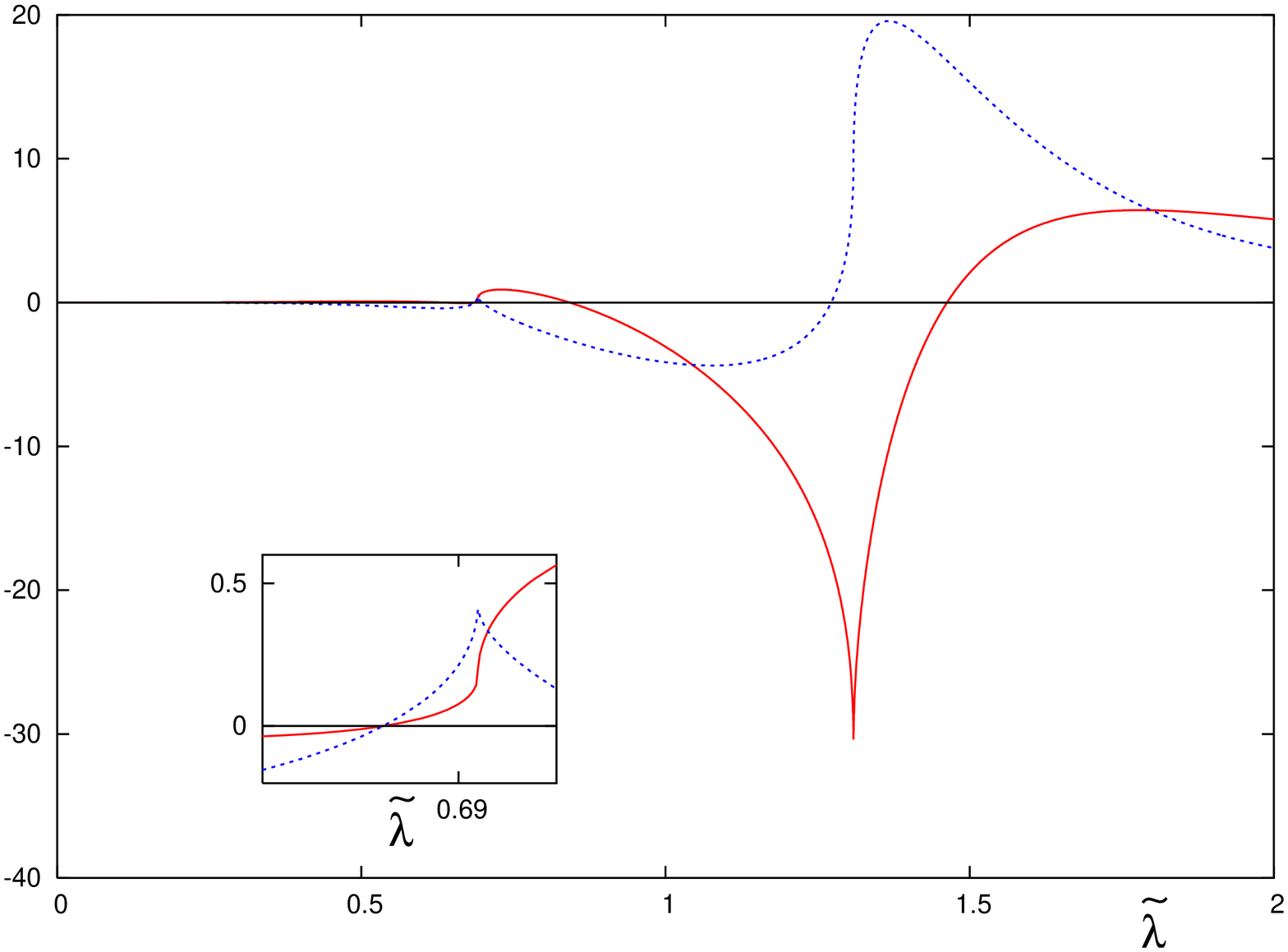}
   \caption{\emph{Real (continuous red line) and imaginary (dashed
       blue line) part of
   the Evans function $E(\lambda)$ as function of the spectral
   parameter~$\lambda$ with $\mbox{\rm off}=0$, for $V=0.9$ and $\theta=0.4\pi$ 
   \newline 
   (a) The integrable case $\rho=0$, The inset is a
   blow-up of the Evans function near the edge of the lower branch
   ($\lambda_-$) of
   the continuous spectrum. \newline (b): $\rho=0.1$. The inset is a
   blow-up of the Evans function and shows the discrete eigenvalue
   where $E(\lambda)=0$. The endpoints of the continuous spectrum occur at
   the cusps of the Evans function.}}
\label{fig0a}
\end{figure}
In the integrable case ($\rho=0$), the linearised massive Thirring
model~\eqref{MTM} has only neutral and continuous
eigenvalues~\cite{kaup961,kaup962} and the solitary wave is stable for
any $0<\theta<\pi$.  The Evans function has zeros at the end points of
the branches of continuous spectrum if $\rho=0$. In
Figure~\ref{fig0a}a, we used our algorithm to calculate the Evans
function for $\rho=0$, while the spectral parameter$\lambda$ is on the
imaginary axis (\mbox{\rm off}=0). This illustrates the zeros of the Evans
function at the end points of the branches of continuous spectrum.  As
follows from Section~\ref{sec.as}, the edges of the continuous
spectrum are located at $\lambda_+=i(|\Omega|+1)$ (upper branch) and
at $\lambda_-=i(1-|\Omega|)$ (lower branch).  
At these endpoints, the real or the imaginary part of the Evans
function exibits a cusp, as can be seen in Figure~\ref{fig0a}a. 
These cusps illustrate the non-analyticity of the Evans function at
those points. 
Note that the cusps and the associated non-analyticity of
the Evans function at the endpoints of the
continuous spectrum occur for all $\rho> 0$ and all
$0<\theta<\pi$, as can be seen from the following figures.

For $0<\theta<\pi/2$ and $\rho>0$, the linearised perturbed massive
Thirring model~\eqref{MTM} has a discrete eigenvalue, which is located
on the imaginary axis in the gap between the branches of continuous
spectrum, see results in~\cite{bpz98, kapitula02}. This eigenvalue has
bifurcated at $\rho=0$ from the lower end point ($\lambda_-$) of the
continuous spectrum.
Our algorithm can follow this discrete eigenvalue on the imaginary
axis, as illustrated in Figure~\ref{fig0a}b. The
edges of the two continuous branches corresponding to positive and
negative energy states in the massive Thirring model (\ref{MTM}) are
located on the imaginary axis at $\lambda_+=i(|\Omega|+1)$ (upper
branch) and at
$\lambda_-=i(1-|\Omega|)$ (lower branch) (see Section~\ref{sec.as}
). Here the discrete eigenvalue which lies on the imaginary axis is
clearly below the edge of the lower branch ($\lambda_-$) of the
continuous spectrum. In the gap between the branches of continuous
spectrum, the Evans function is still analytic, since there is still a
2:2 split of the dimensions of the stable and unstable manifolds.

\begin{figure}[bth]
\quad
\includegraphics[width=.93\textwidth]{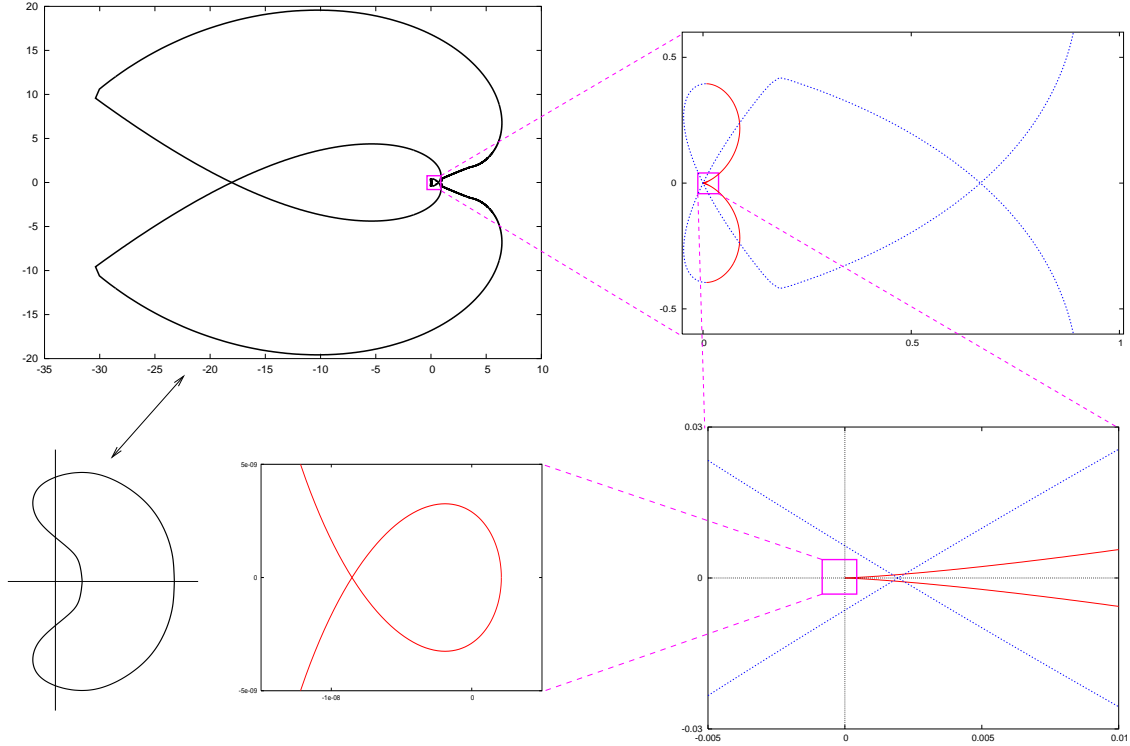}
  \caption{\emph{The real versus imaginary parts of the Evans function
  $E(\lambda)$ for the non-integrable case $\rho=0.1$, $V=0.9$ and
  $\theta=0.4\pi$. The spectral parameter $\lambda$ varies parallel to
  the imaginary axis with a small offset $\mbox{\rm off}=10^{-11}$.
  Going round clockwise from the upper left corner:\newline 
  (a): Overview of the Evans function.\newline 
  (b): First zoom into the area near $E=0$. The colours are used to
  help identifying the lines in the next picture.\newline 
  (c): Next zoom into this area. This picture shows that the
  two little loops cross right of the zero point. 
For visualisation  purposes, we used $\mbox{\rm off}=5\cdot 10^{-5}$. 
 At $\mbox{\rm off}=0$ the crossing of the loops is located exactly at the
 origin, indicating the discrete eigenvalue on the imaginary axis. 
 The slight offset $\mbox{\rm off}=5\cdot 10^{-5}$ moves the crossing
   of the loops to the right of the origin. 
\newline
  (d): Final zoom into the area near $E=0$ (note how small the scale
  is). We have increased the offset to $\mbox{\rm off}=5\cdot 10^{-3}$ to
  avoid problems with the smallness of the Evans function near
  $\lambda=0$.\newline
  (e): A topologically  equivalent sketch of the Evans function.
}}\label{fig0d}
\end{figure}
For $0<\theta<\pi/2$, the solitary waves are known to be stable even
for non-zero $\rho$ \cite{bpz98, kapitula02}. 
In Figure~\ref{fig0d} we
show the Evans function for such a stable case.
To determine the winding number, we need to zoom into the
neighbourhood of $E=0$. The behaviour near $E=0$
(Figure~\ref{fig0d}b--d)
and the global behaviour of the Evans function ({Figure~\ref{fig0d}a) 
convey that the Evans function is topologically equivalent to a loop
which does not contain the origin (Figure~\ref{fig0d}e). Hence the
winding number is zero, confirming stability,

For $\pi/2<\theta<\pi$, instabilities may arise for non-zero
$\rho$. These instabilities are of an oscillatory nature in the sense
that discrete eigenvalues leave the imaginary axis into the complex
plane \cite{bpz98,kapitula02}. The onset of the first instability is
when the discrete eigenvalue in the gap of the continuous spectrum
merges with the 
edge of the lower branch ($\lambda_-$) of the continuous spectrum. In the
$\theta$-$\rho$ bifurcation plane, the instability curve starts at
$\rho=0$ and $\theta=0.5$
initially proportional to $\sqrt\rho$~\cite{bpz98,kapitula02}. 
\begin{figure}[tb]
  \begin{center}
  \includegraphics[width=.3\textwidth]{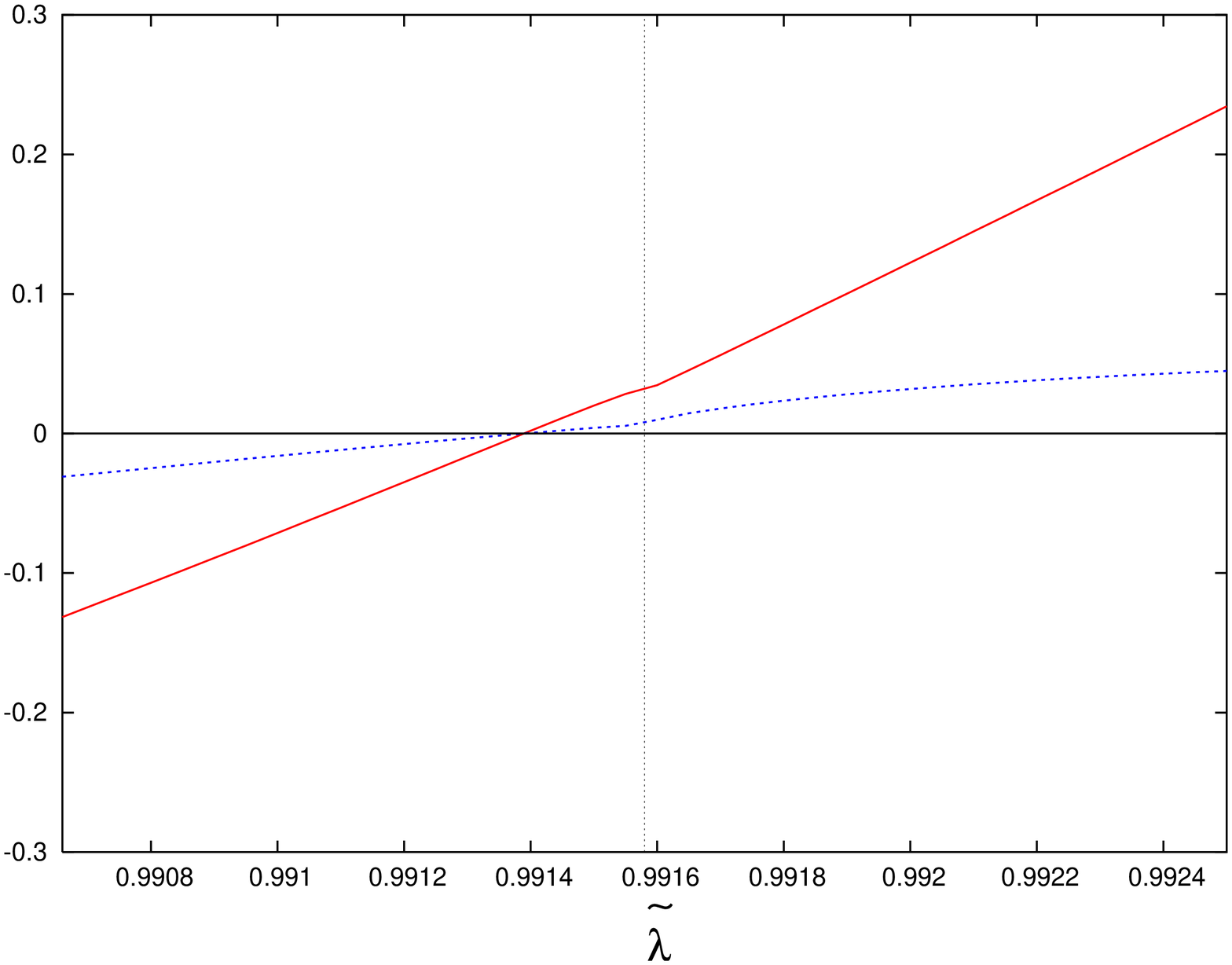}\qquad
  \includegraphics[width=.3\textwidth]{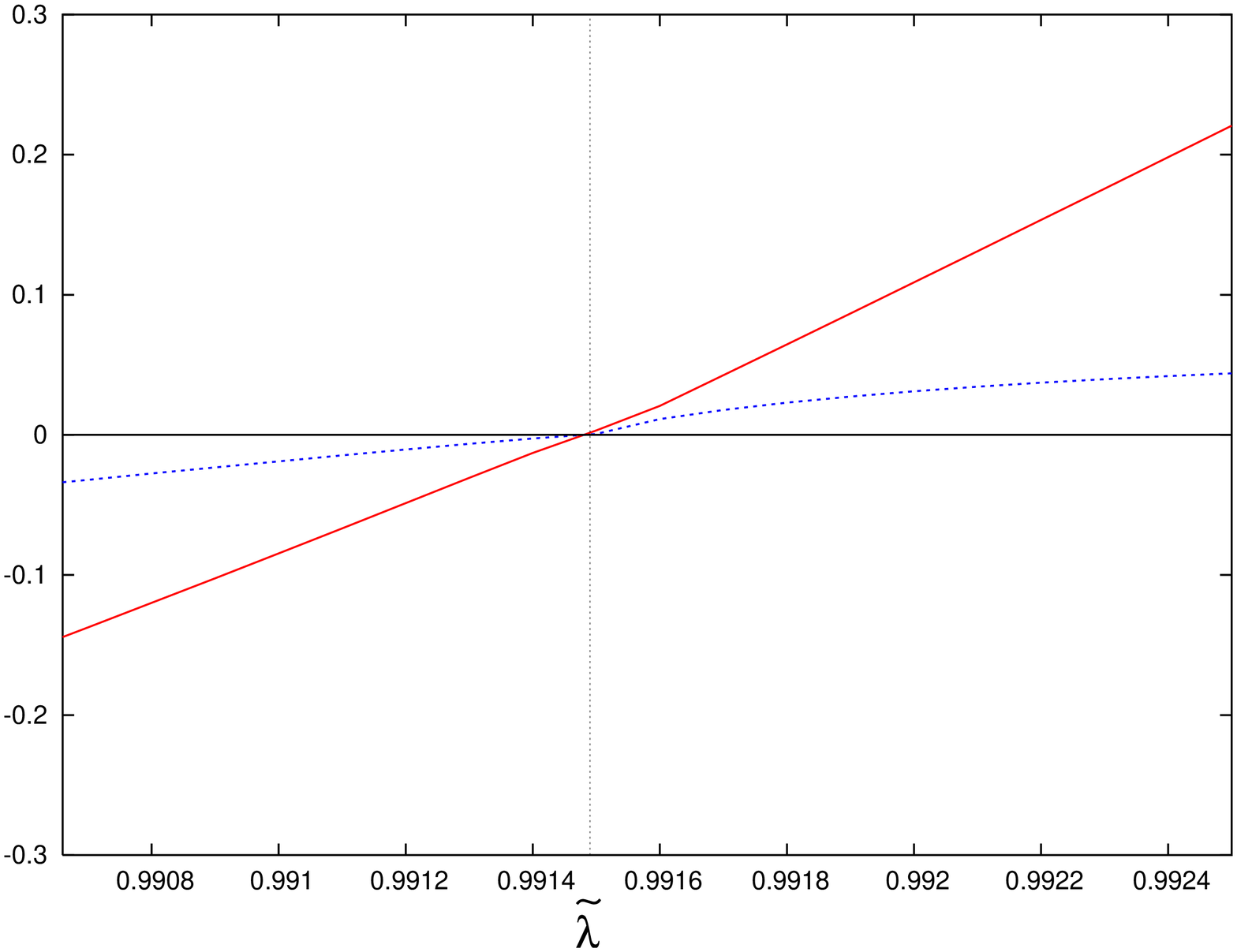}\qquad
  \includegraphics[width=.3\textwidth]{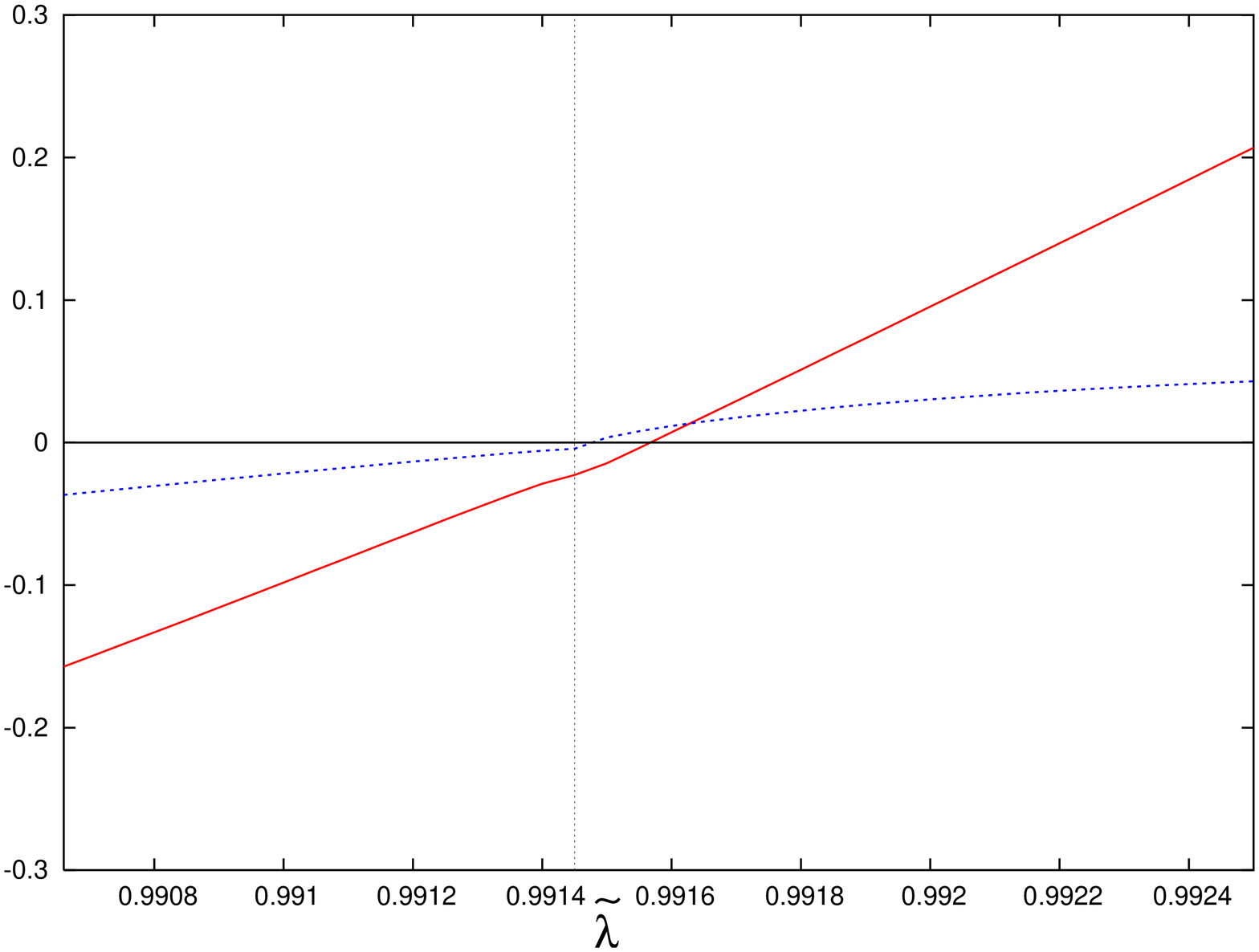}
 \end{center}
 \caption{\emph{The onset of the primary instability for $\rho=0.1$ and
     $V=0.9$ (the offset $\mbox{\rm off}=0$). Plotted are the real part
     (continuous red line) and imaginary part (dashed blue line) of
     $E(\lambda)$ versus $\lambda$. The vertical dotted line indicates
     the edge of the lower branch ($\lambda_-$) of
     the continuous spectrum. In the left picture with
     $\theta=0.50268\pi$, the eigenvalue is still located on the
     imaginary axis. In the middle picture with $\theta = 0.50270\pi$,
     the eigenvalue has just merged with the end point of the lower
     branch ($\lambda_-$) of the continuous spectrum, giving a zero of
     the Evans function on the imaginary axis at the end of the
     continuous spectrum. In the right picture
     with $\theta=0.50272\pi$, there is no eigenvalue on the imaginary
     axis anymore.
}}
\label{fig3} 
\end{figure}
%
\begin{figure}[htb]
\qquad  \includegraphics[angle=-90,width=.47\textwidth]{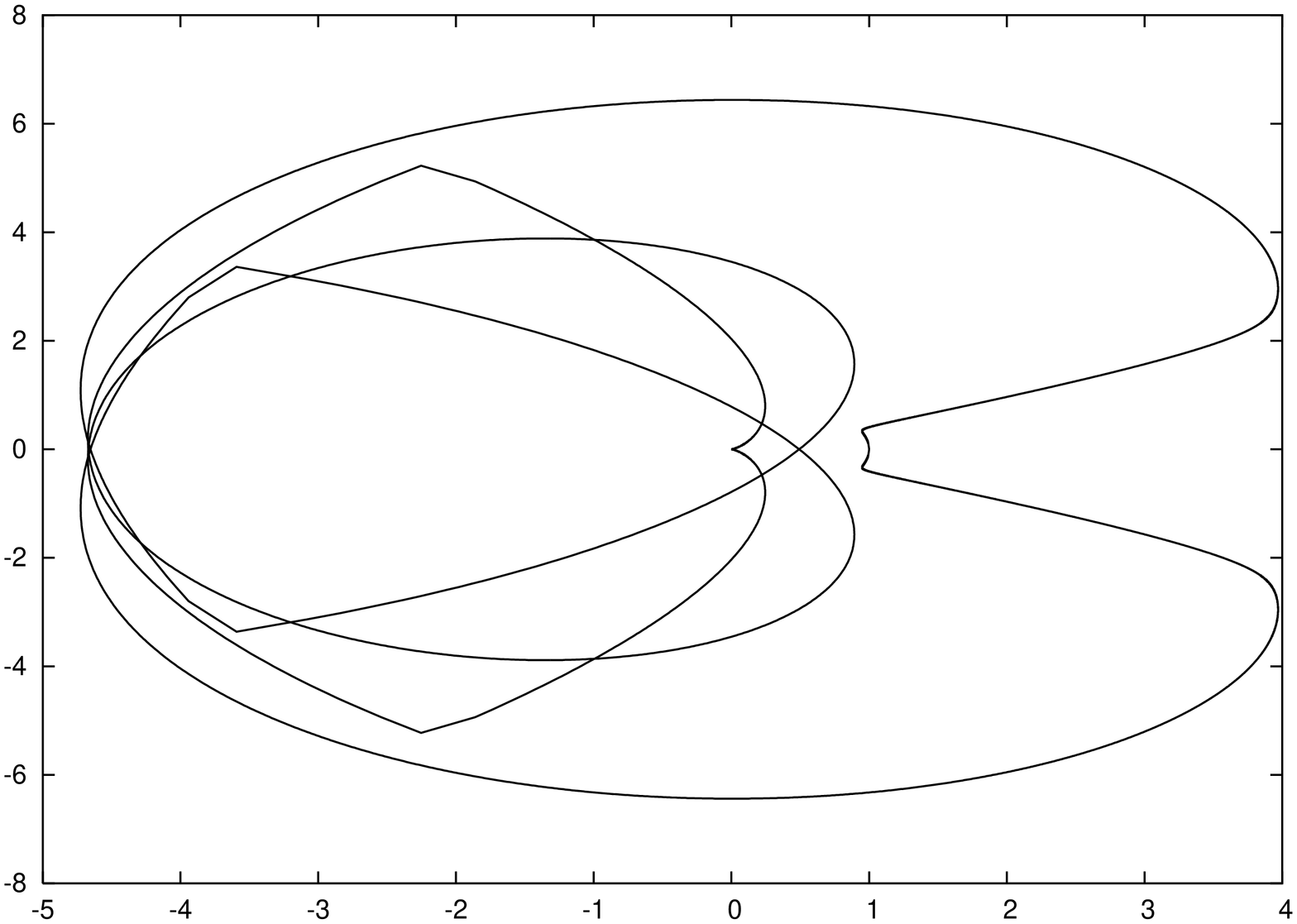}
\vskip -5.2cm \hskip 10cm
  \includegraphics[angle=0,width=.3\textwidth]{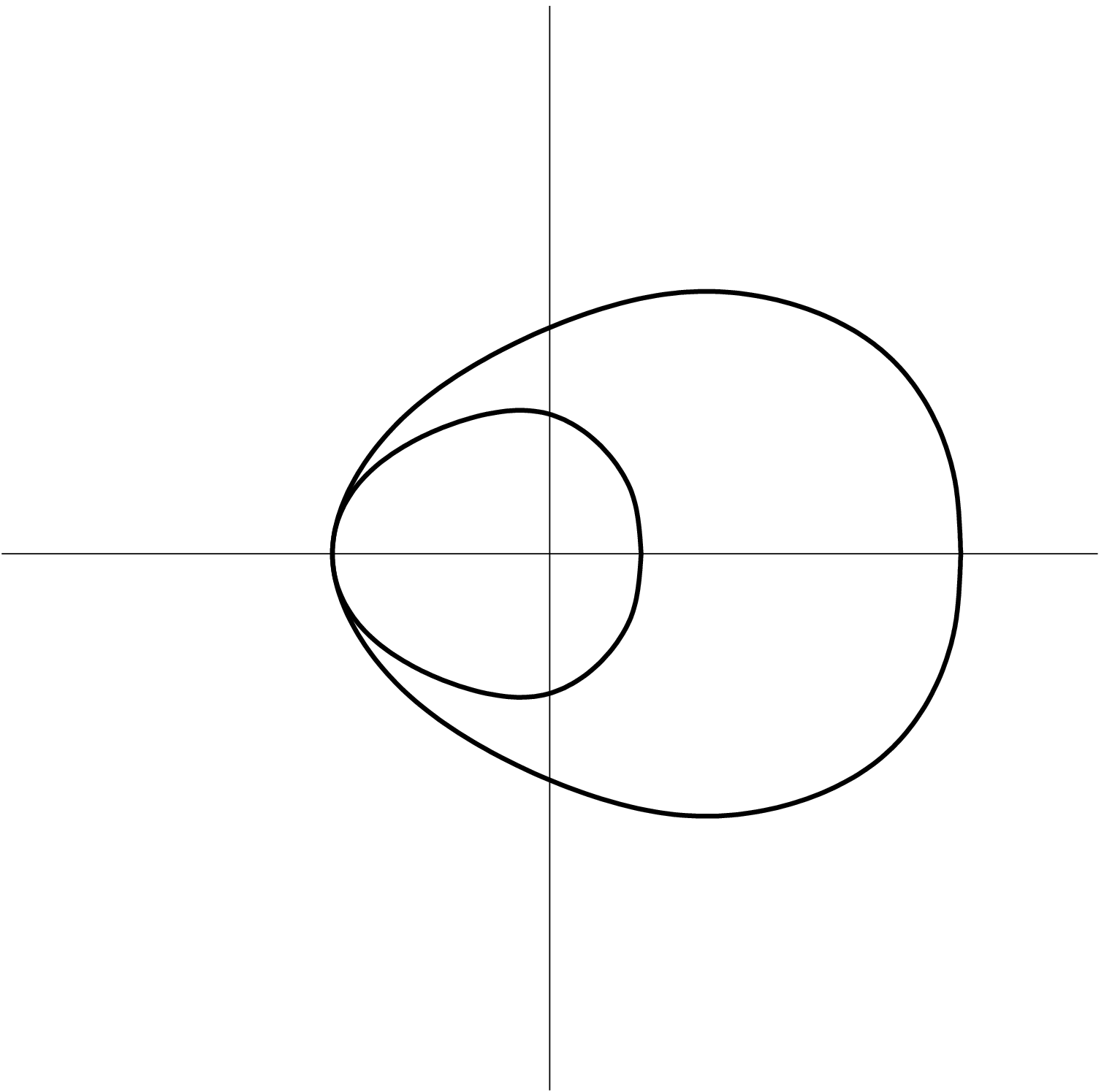}
  \caption{\emph{(a): 
The real versus imaginary parts of the Evans function
  $E(\lambda)$ for $\rho=0.1$, $V=0.9$ and
  $\theta=0.6\pi$. 
The spectral parameter $\lambda$ varies parallel to
the imaginary axis with $\mbox{\rm off}=5\cdot10^{-14}$.
\newline (b): The Evans 
  function in Figure~\ref{fig2}a is topologically equivalent to this
  sketch of $E(\lambda)$. The winding number is clearly~$2$,
  confirming that there is one pair of unstable eigenvalues for these
  parameter values.}}  \label{fig2}
\end{figure}
In Figure~\ref{fig3}, the onset of the instability is illustrated at
$\rho=0.1$ and $V=0.9$. 
%
At the bifurcation point the eigenvalue detaches from the imaginary
axis at the edge of the lower branch ($\lambda_-$) of the continuous spectrum and
leaves into the complex plane. 
This can be explored by looking at the
winding number of the Evans function. Figure~\ref{fig2} shows the real
and imaginary parts of the Evans function when the instability has
well occurred (at $\rho=0.1$, $V=0.9$ and $\theta=0.6\pi$). The winding
number is~2, confirming that a pair of unstable eigenvalues is
present.
%
%

If we now decrease $\rho$ (with $\theta$ fixed) from the above
described primary bifurcation of oscillatory instability, the primary
unstable eigenvalue will merge with the edge of the upper branch
($\lambda_+$) of the continuous spectrum. This is illustrated in
Figure~\ref{fig4}. If $\rho$ is decreased below this bifurcation, the
solitary wave is stable again. We see clearly the collision of the
eigenvalue with the edge of the upper branch ($\lambda_+$) of the
continuous spectrum.
In~\cite{kapitula02}, a slightly different perturbation is studied and
it is proved analytically that in this case two bifurcation curves
originate from $(\theta,\rho)=(0.5\pi,0)$. However, it is not clear
what the relation with the branches of continuous spectrum is away
from the bifurcation point. Our calculations suggest that for our
perturbation there are again two bifurcation curves, one curve is
related to a bifurcation from the edge of the lower branch
($\lambda_-$) and the other curve to a bifurcation from the edge of
the upper branch ($\lambda_+$) and the primary unstable eigenvalue
goes between these two branches.  This seems a new observation, which
has not been recognised earlier.
   
\begin{figure}[htb]
  \begin{center}
\qquad  \includegraphics[width=.47\textwidth]{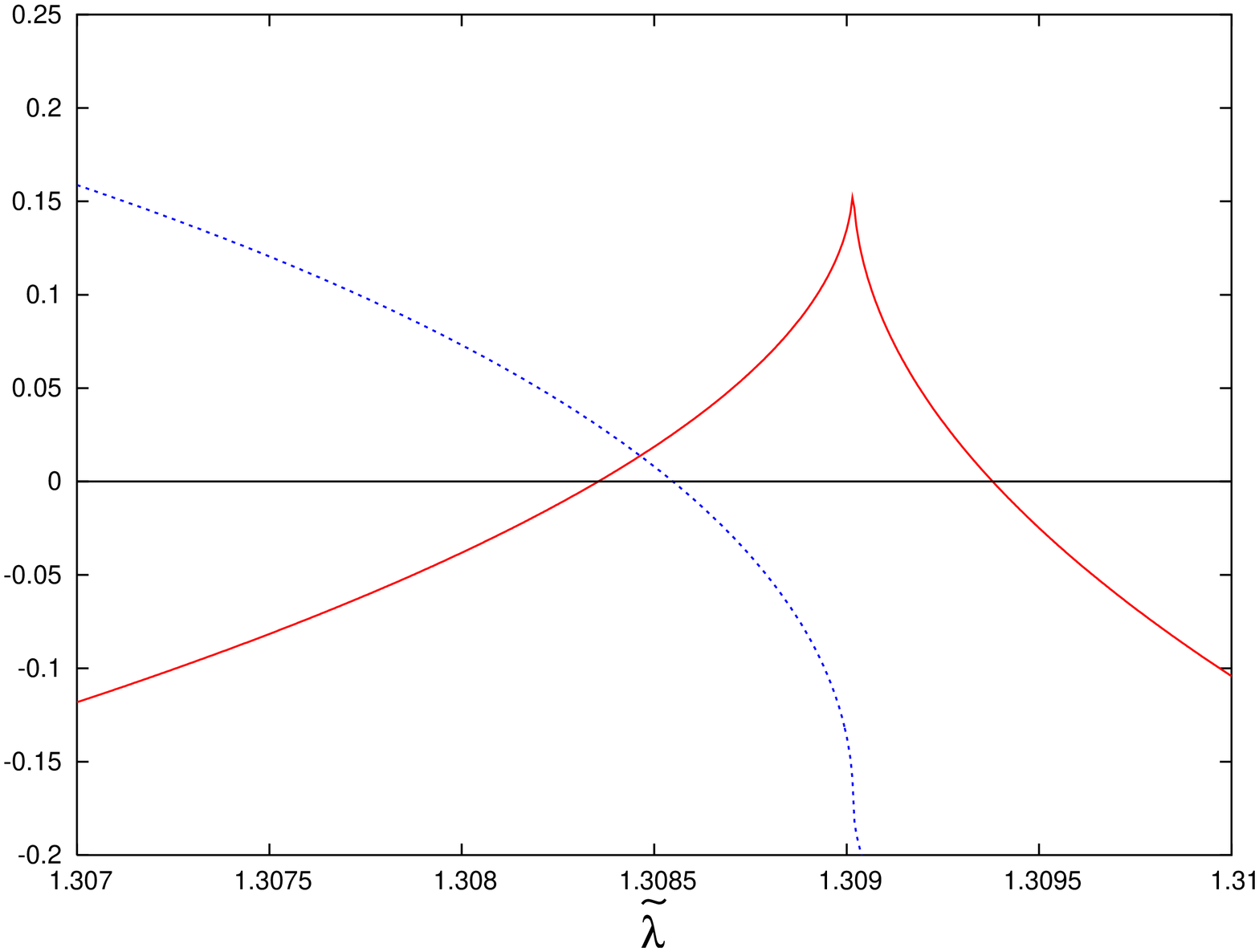}
\includegraphics[width=.47\textwidth]{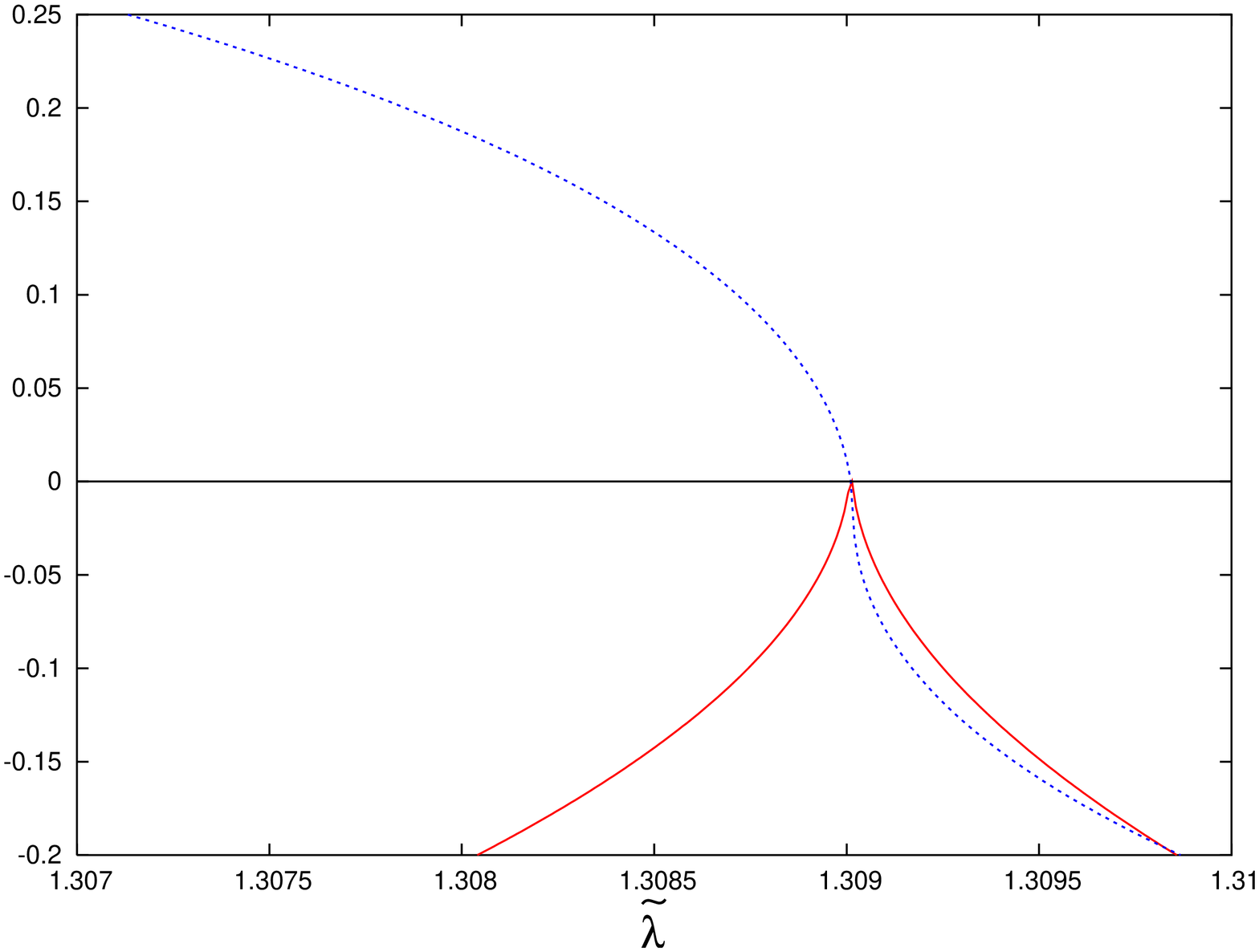}
  \caption{\emph{Real (continuous red line) and imaginary part (dashed
      blue line) of
  the Evans function $E(\lambda)$ as a function of the spectral
  parameter $\lambda$ for $V=0.9$ and $\theta=0.6\pi$ (the offset
  $\mbox{\rm off}=5\cdot 10^{-14}$). (a): Above the
  bifurcation point at $\rho=0.002$. (b): At the bifurcation point
  $\rho=0.0001$.}} \label{fig4} \end{center}
\end{figure}

Figures~\ref{fig3} and~\ref{fig4} illustrate that our algorithm and
the Evans function allow us to exactly follow the eigenvalues in the
course of the bifurcation.  By looking at the real and imaginary part
of the Evans function the value of lambda which corresponds to a root
of the Evans function can be located. This allowed us to answer the
question from where on the imaginary axis the eigenvalues detach for the
primary instability.

It has been found analytically in \cite{bpz98} that
secondary bifurcations can occur for suitable values of $\rho$ and
$\theta$ (we assume $V$ fixed here for simplicity). Here a second pair
of eigenvalues detach from the continuous spectrum into the complex
plane.  Again, analyticity of the Evans function allows us to compute
the number of unstable eigenvalues by determing the winding number. In
Figure~\ref{fig5} we show that the winding number equals~$4$ when
$\rho=0.1$, $V=0.9$ and $\theta=0.9\pi$, indicating that a new pair of
eigenvalues have detached. 
%
\begin{figure}[htb]
  \qquad\includegraphics[angle=-90,width=.47\textwidth]{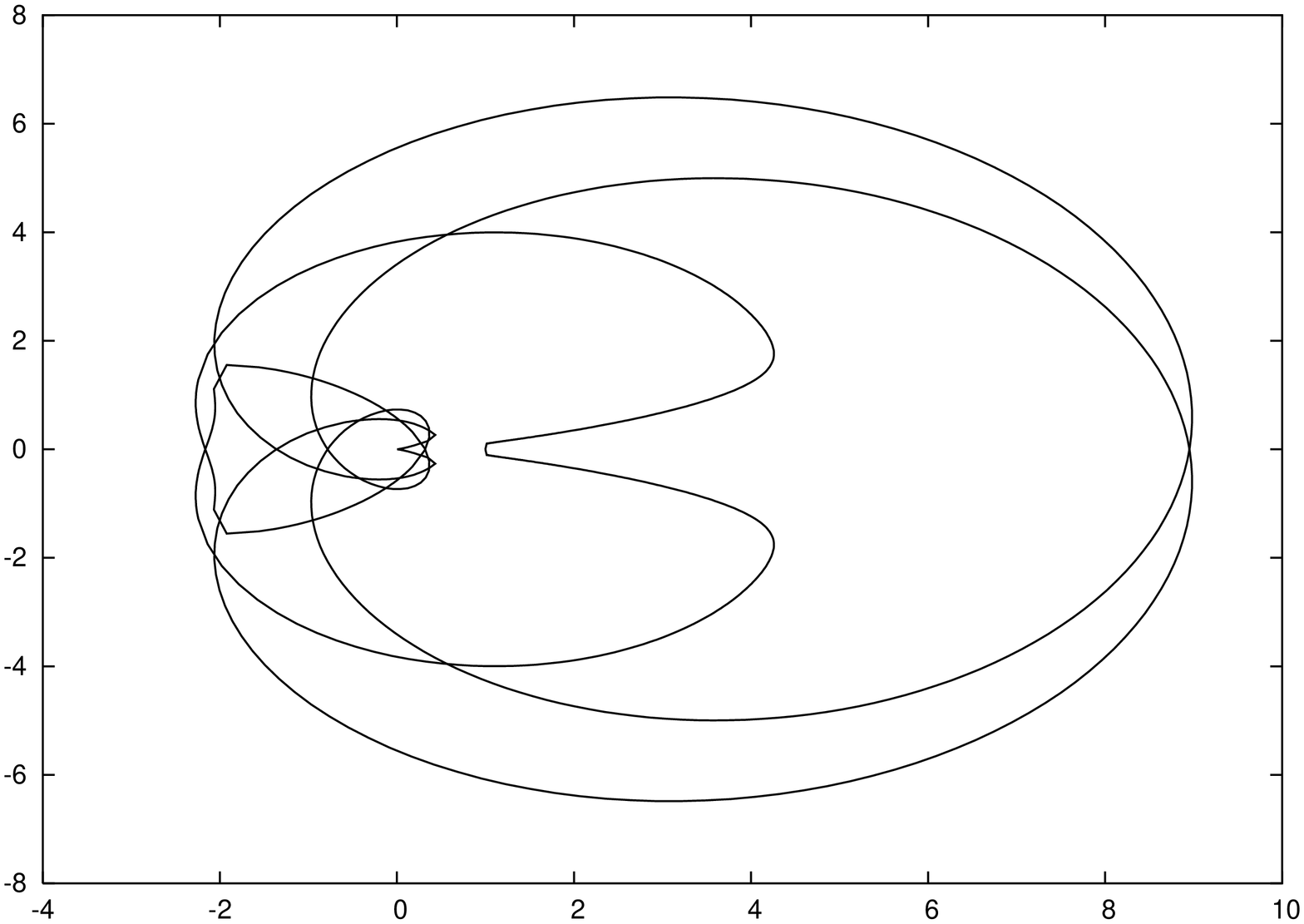}
\vskip -5.2cm \hskip 10cm
  \includegraphics[angle=0,width=.3\textwidth]{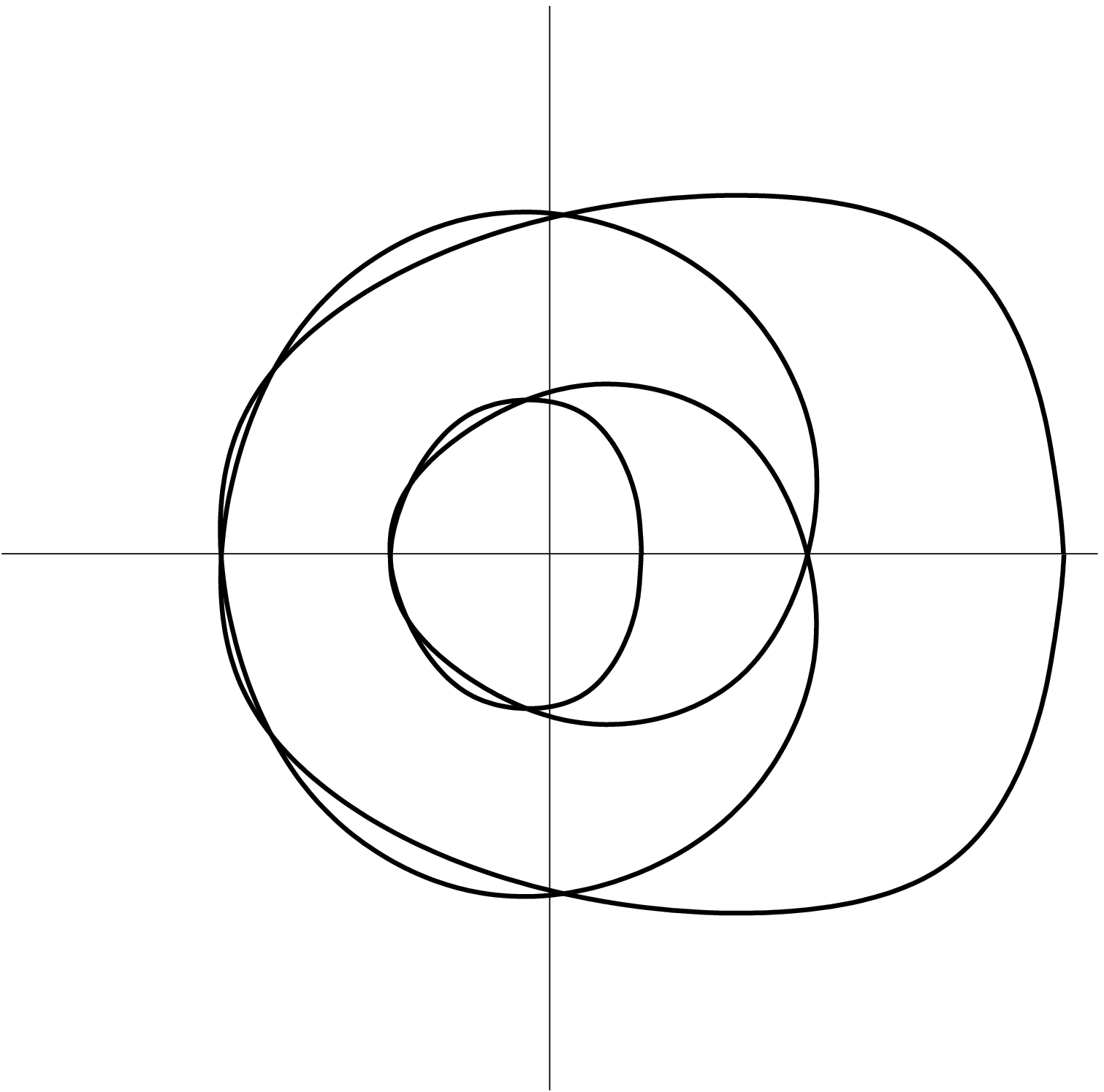}
  \caption{\emph{(a): 
The real versus imaginary parts of the Evans function
  $E(\lambda)$ for $\rho=0.1$, $V=0.9$ and
  $\theta=0.9\pi$. The spectral parameter $\lambda$ varies parallel to
  the imaginary axis with $\mbox{\rm off}=5\cdot10^{-14}$.
\newline
(b): The Evans function in (a) is topologically
  equivalent to 
  this picture. The winding number is clearly $4$, confirming that
  there are two pairs of unstable eigenvalues for these parameter values.}} 
  \label{fig5}
\end{figure}
The behaviour near $\lambda=0$ close to the origin of the Evans
function behaves is similar to the one depicted in
Figure~\ref{fig0d}d.  The onset of this secondary instability at
$\rho=0.1$ and $V=0.9$ is shown in Figure~\ref{fig6}.
Figure~\ref{fig6} shows the behaviour of the Evans function shortly
before, at, and shortly after the secondary instability. In order to
see the secondary instability and the emergence of the eigenvalue, one
needs to look at the behaviour near $E=0$. Note that the secondary
instability is hard to observe in the full Evans function (left
pictures in Figure~\ref{fig6}).  In the middle picture, the Evans
function has a zero, just after the edge of the upper branch
($\lambda_+$) of the continuous spectrum, but not at the edge of this
branch.  Hence for $\rho>0$, the eigenvalue comes really from within
the continuous spectrum, not out of one the endpoints of the branches
of the continuous spectrum. Similar behaviour has been described by
\textsc{Sandstede \& Scheel}~\cite{sandstede00}.



\bigskip 
The results in this section show that our algorithm using the Evans
function approach exercised on the wedge space provides a good and
reliable diagnostic tool for the accurate detection of oscillatory
instabilities. The analyticity of the Evans function allows for
detection of complex eigenvalues by assuring the necessary conditions
for the application of Cauchy's principle. Our algorithm is able to
follow the location of the discrete eigenvalues in the course of their
bifurcations. The Evans function does not involve a discretization of
the spectrum of the eigenvalue problem and hence does not fracture the
continuous spectrum. It is therefore free of spurious unstable
eigenvalues. Moreover, the exterior algebra is a platform for
numerically stable shooting. As already stated, the method is not
restricted to the particular coupled mode model (\ref{MTM}) but 
can be applied to other problems as well.
 \begin{figure}[htb]
   \begin{center}
    \includegraphics[angle=-90,width=.47\textwidth]{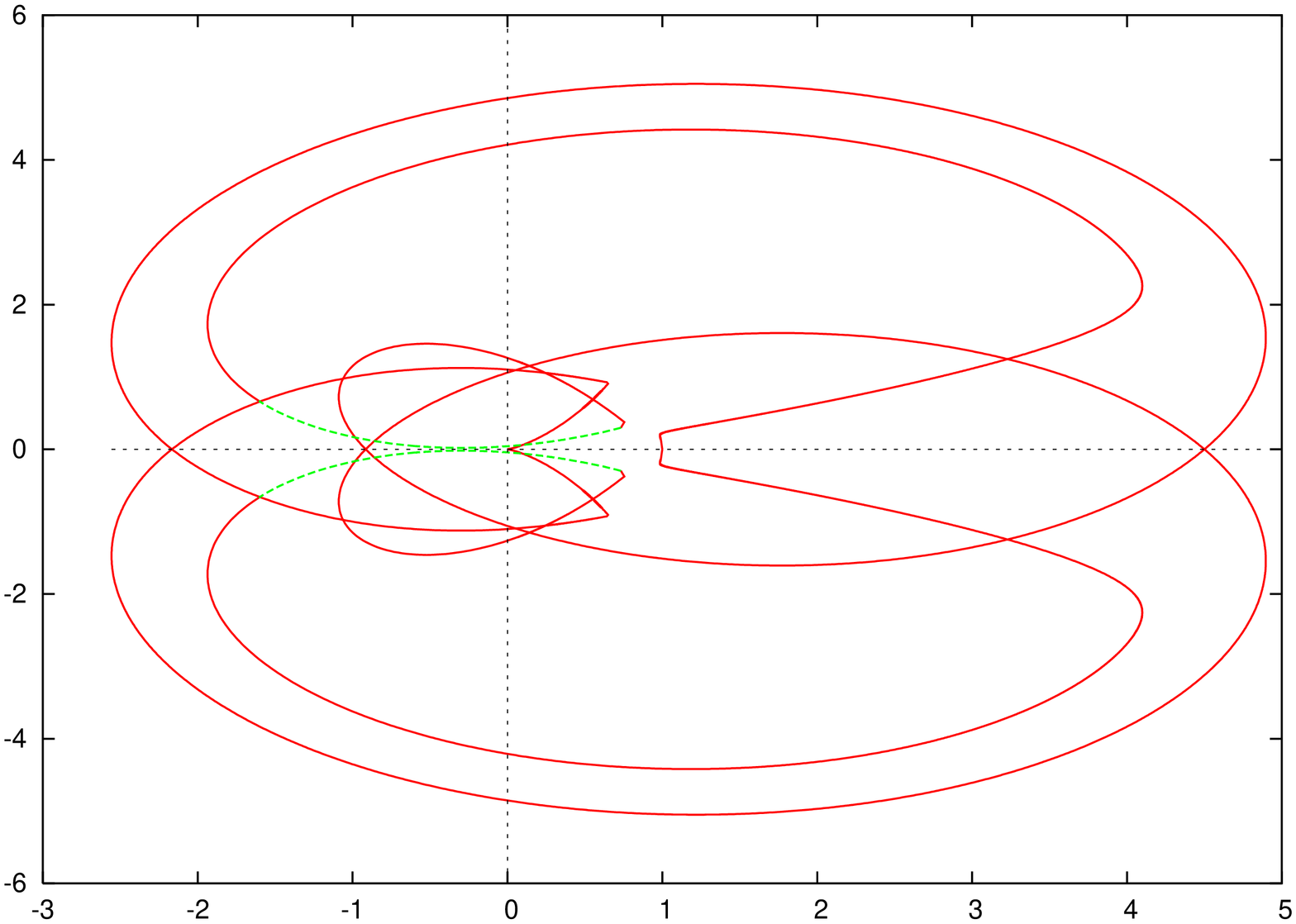}\quad
    \includegraphics[angle=-90,width=.47\textwidth]{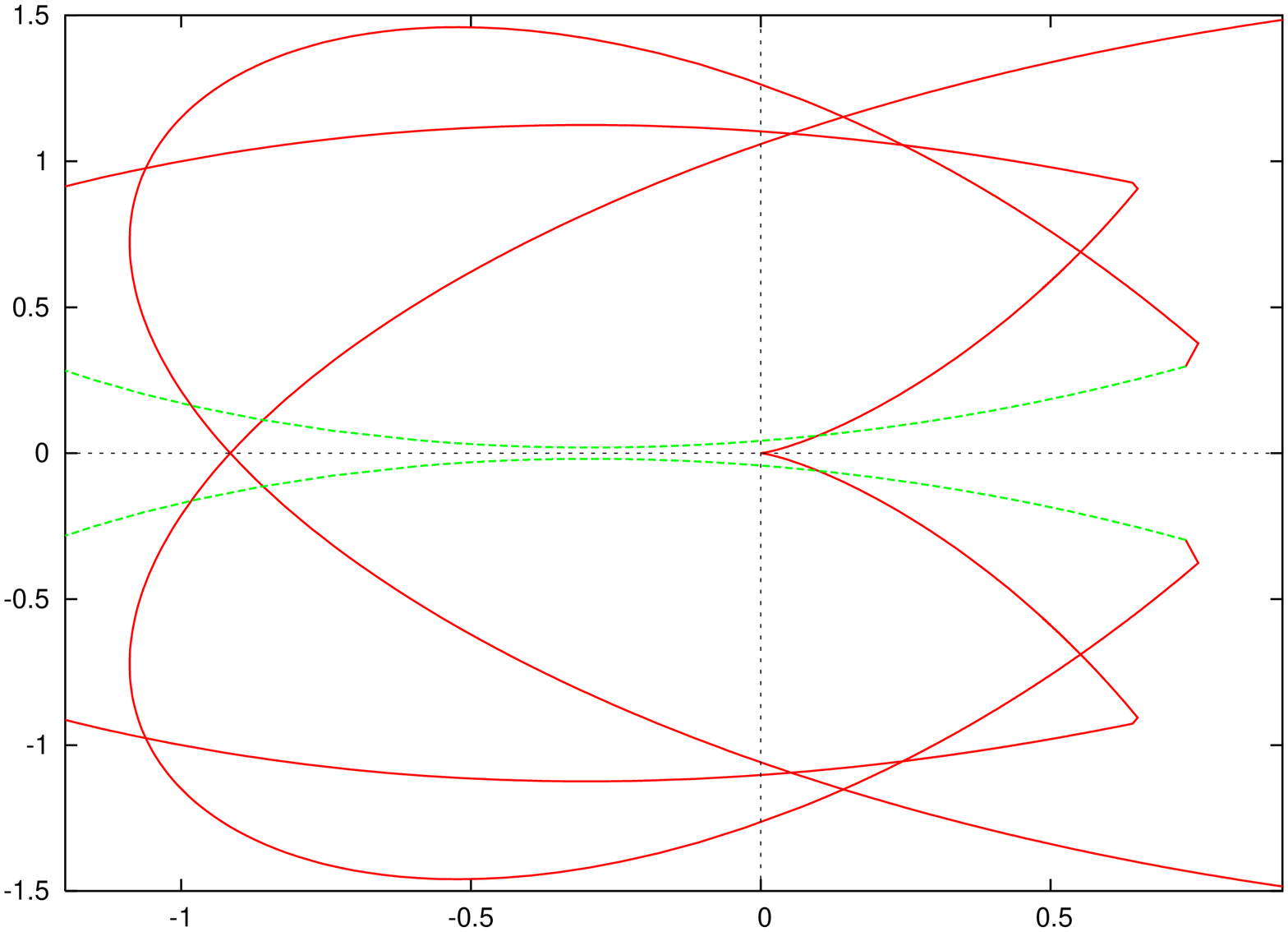}

    \includegraphics[angle=-90,width=.47\textwidth]{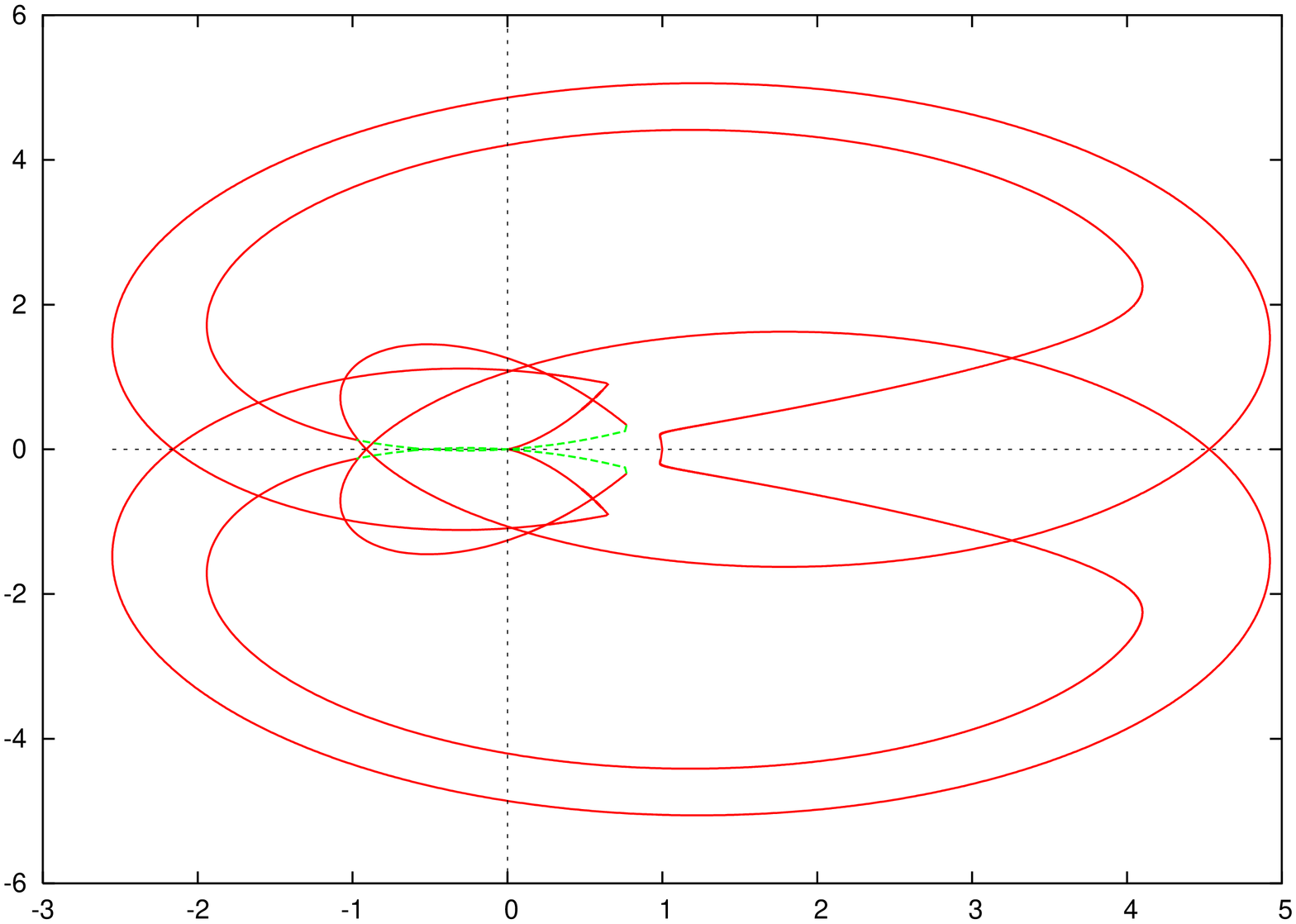}\quad
    \includegraphics[angle=-90,width=.47\textwidth]{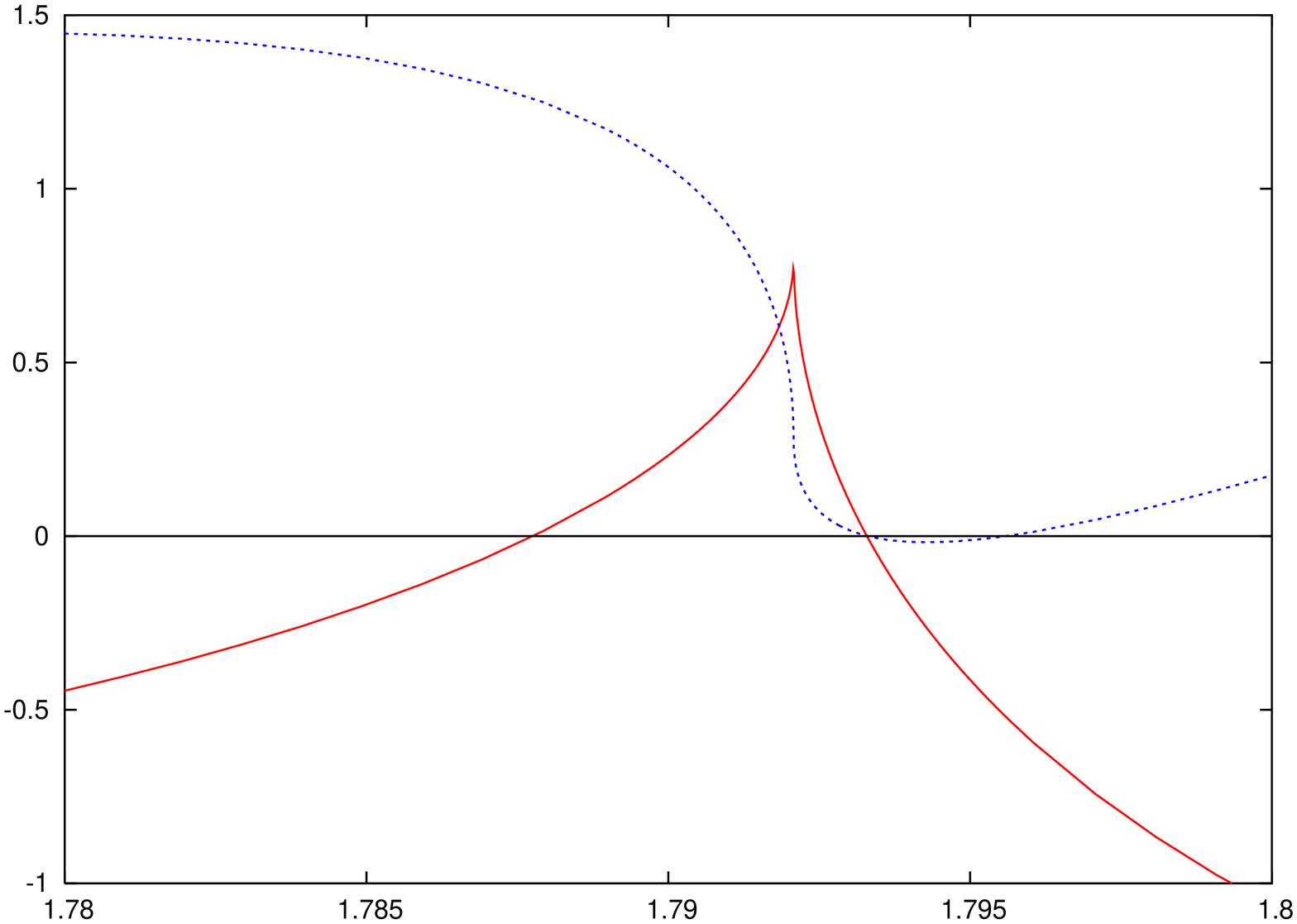}

    \includegraphics[angle=-90,width=.47\textwidth]{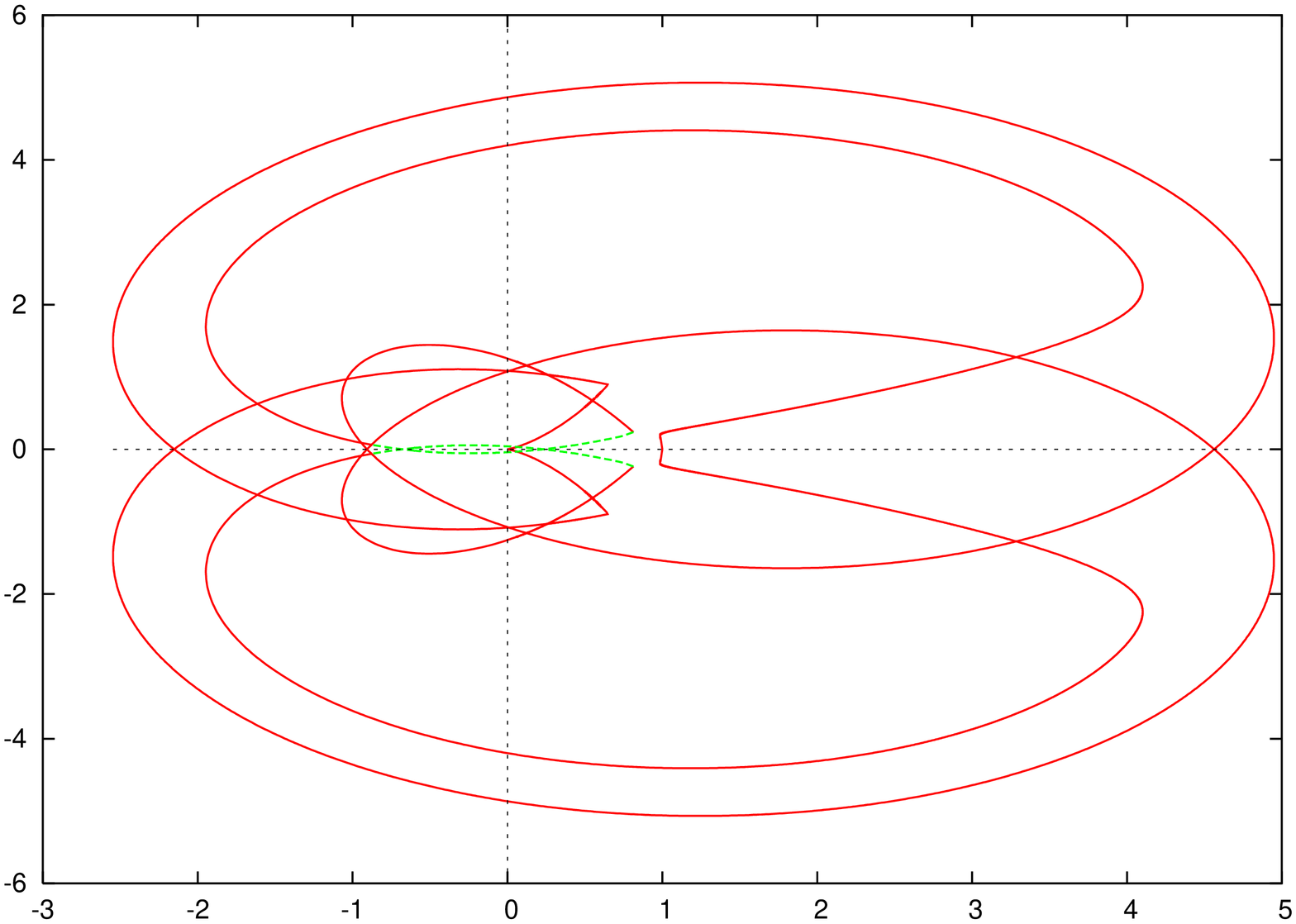}\quad
    \includegraphics[angle=-90,width=.47\textwidth]{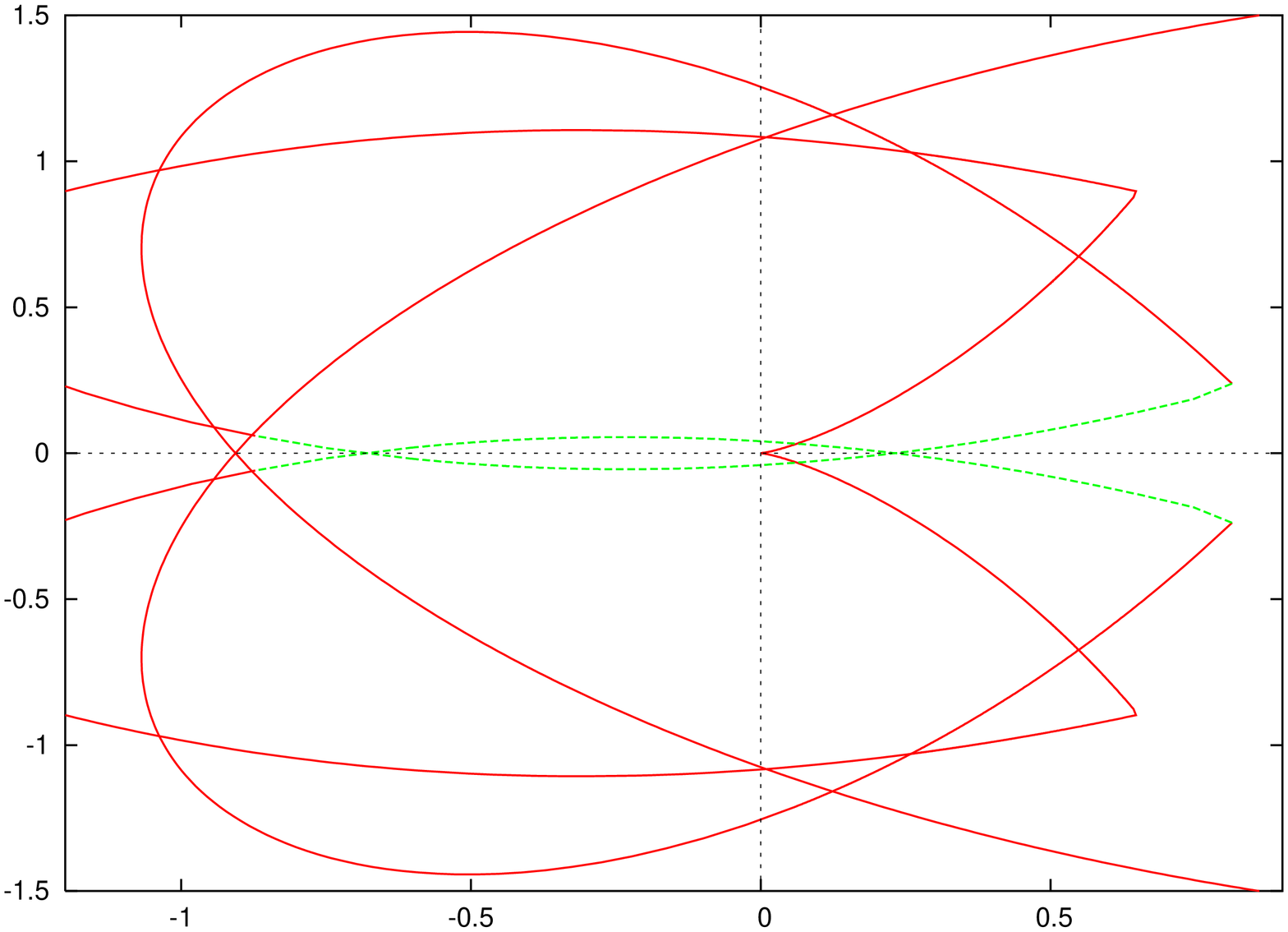}
   \end{center}
    \caption{\emph{
The real versus imaginary parts of the Evans function
  $E(\lambda)$ for $\rho=0.1$, $V=0.9$ and $\theta$ varying. The
  spectral parameter $\lambda$ varies on 
  the imaginary axis ($\mbox{\rm off}=0$).
The green line helps to locate the emergence of the second eigenvalue.
\newline 
        (a):
        Just before the onset of the secondary instability at
        $\theta=0.790$. The right picture zooms in at the neighbourhood
        of $E=0$.
\newline 
        (b): At the onset of instability at
        $\theta=0.791$. The right picture shows the real (continuous
        red line) and imaginary (dashed blue line) parts of the Evans
        function as function of $\lambda$.
        This shows clearly that the zero of the Evans function occurs
        just after the edge of the upper branch ($\lambda_+$) of
        continuous spectrum (i.e., the cusp), hence inside the
        continuous spectrum and not at the edge.
        \newline 
        (c): Just after the onset of the secondary instability
        at $\theta=0.792$. The right picture zooms into the
        neighbourhood of $E=0$. The loops with the green line now
        include the origin ($E=0$), increasing the winding number by~$2$
        and illustrating that an additional instability has occurred.
}}
\label{fig6}
 \end{figure}


\section*{Acknowledgements}\label{ack}
We would like to thank Thomas Bridges and Dmitry Pelinovsky for
valuable discussions. Gianne Derks' research was partially supported by
a European Commission Grant, contract number HPRN-CT-2000-00113, for
the Research Training Network {\it Mechanics and Symmetry in Europe\/}
(MASIE), {\tt http://www.ma.umist.ac.uk/jm/MASIE/}

%
%

\end{document}